\documentclass[11pt,twocolumn]{aastex631}

\usepackage{ mathrsfs }

\begin{document}

\title{A Tentative Detection of a Point Source in the Disk Gap of HD 100546 with VLT/SPHERE-IRDIS Sparse Aperture Masking Interferometry \footnote{ Based on observations collected at the European Southern Observatory under ESO programmes 2100.C-5052(B), 105.2067.001}}

\author[0000-0001-9582-4261]{Dori Blakely}
\affiliation{Department of Physics and Astronomy, University of Victoria, 3800 Finnerty Road, Victoria, BC, V8P 5C2, Canada}
\affiliation{NRC Herzberg Astronomy and Astrophysics, 5071 West Saanich Road, Victoria, BC, V9E 2E7, Canada}

\author[0000-0002-6773-459X]{Doug Johnstone}
\affiliation{NRC Herzberg Astronomy and Astrophysics,
5071 West Saanich Road,
Victoria, BC, V9E 2E7, Canada}
\affiliation{Department of Physics and Astronomy, University of Victoria, 3800 Finnerty Road, Victoria, BC, V8P 5C2, Canada}

\author[0000-0002-5823-3072]{Tomas Stolker}
\affiliation{Leiden Observatory, Leiden University, Einsteinweg 55, 2333 CC Leiden, The Netherlands}

\author[0000-0002-7695-7605]{Myriam Benisty}
\affiliation{Max-Planck Institute for Astronomy (MPIA), Königstuhl 17, 69117 Heidelberg, Germany}

\author[0000-0003-2769-0438]{Jens Kammerer}
\affiliation{European Southern Observatory, Karl-Schwarzschild-Straße 2, 85748 Garching, Germany}

\author[0000-0001-5898-2420]{Brodie J. Norfolk}
\affiliation{School of Physics and Astronomy, Monash University, Vic 3800, Australia}

\author[0000-0001-5684-4593]{William Thompson}
\affiliation{NRC Herzberg Astronomy and Astrophysics,
5071 West Saanich Road,
Victoria, BC, V9E 2E7, Canada}

\author{Jean-Philippe Berger}
\affiliation{Universit\'{e} Grenoble Alpes, CNRS, IPAG, 38000 Grenoble, France}



\begin{abstract}
We re-analyze VLT/SPHERE-IRDIS \textit{K} and \textit{H}-band sparse aperture masking interferometry data of the transition disk HD 100546 observed in 2018 and 2021, respectively. We fit geometrical models to the closure phases extracted from both datasets. We compare three model classes: a forward scattering disk, a forward scattering disk plus an arbitrary asymmetric disk feature and a forward scattering disk plus an unresolved point source in the disk-gap. {We find that the forward scattering disk plus point source model is the best representation of the data.}
We find that this point source candidate moved from a position of {sep.\ = $39.9^{+2.8}_{-3.3}$ mas, P.A. = $124.1^{+1.0}_{-1.0}$ degrees to a sep.\ = $50.0^{+1.0}_{-1.0}$ mas, P.A. = $106.4^{+1.4}_{-1.4}$} degrees between 2018 and 2021. {Both of these positions are well within the $\sim$13 au ($\sim$120 mas) disk-gap, favouring the point source interpretation}. We explore the orbital parameter space that is consistent with the measured relative astrometry. We find orbits either with a similar orientation to the outer disk, with a high eccentricity $e \gtrapprox 0.65$, or orbits with a large relative inclination ($\sim$60 degrees) to the outer disk, and any eccentricity.
Despite the significance of the observed {point-source} signal, follow-up observations will be necessary to conclusively determine its nature.

\end{abstract}



\section{Introduction} \label{sec:intro}

\begin{figure*}
\centering
\includegraphics[width=0.85\linewidth]{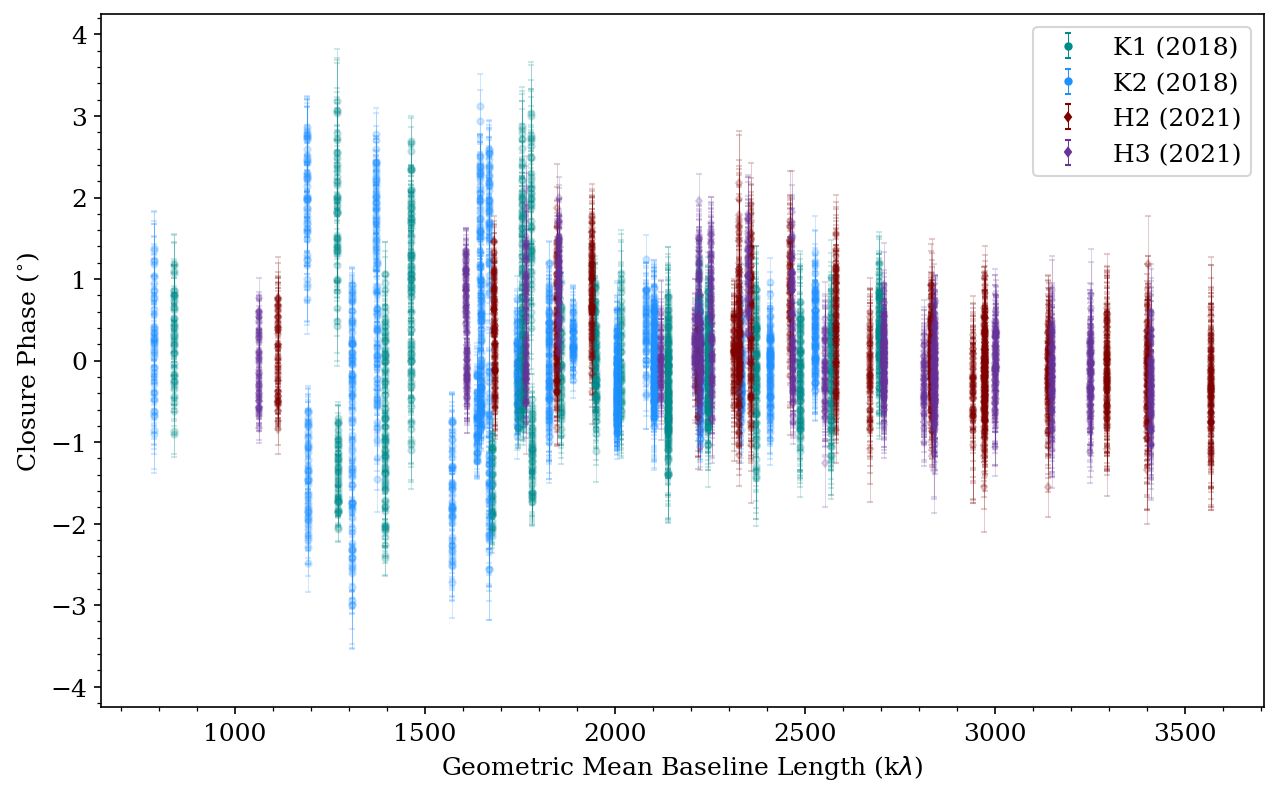}
\caption{{Calibrated closure phases for HD\,100546, extracted from the 2018 2.1 $\mu$m (teal) and 2.3 $\mu$m data (blue) and 2021 1.6 $\mu$m (maroon) and 1.7 $\mu$m data (purple).}}
\label{fig:all_data}
\end{figure*}

The disk around the young star HD 100546 is highly perturbed, with many disk substructures and asymmetries seen in the dust continuum, molecular line emission and near-infrared scattered light \citep[e.g.,][]{2011ApJ...738...23Q,2011A&A...531A...1T,2015MNRAS.453..414W, 2019MNRAS.485..739M, 2021MNRAS.502.5779N, 2021A&A...651A..90F,2023A&A...669A..53B,2023NatAs...7..684K}. These 
{structures}
include a {$\sim$13 au} inner gap, significant deviations from Keplerian motion seen in {CO line} velocity channel maps 
\citep[e.g.\ the Doppler flip;][]{2019ApJ...883L..41C,2020ApJ...889L..24P,2022ApJ...933L...4C}, and spiral arms that are observed in infrared scattered light \citep{2013A&A...560A..20B,2017AJ....153..264F}, as well as in CO line emission \citep{2022ApJ...936L...4N}.

Some of these structures have been hypothesized to be
produced by planetary mass companions. These candidate protoplanets have been either tentatively directly observed but not yet independently confirmed \citep{2013ApJ...766L...1Q,2014ApJ...796L..30C,2015ApJ...814L..27C} or inferred due to unresolved excess CO emission in Keplerian rotation \citep{2019ApJ...883...37B}. 
An additional hypothesis, proposed by \cite{2022ApJ...936L...4N}, is that a yet to be observed eccentric stellar mass companion could be producing many of the observed features. Previous observations, however, have placed some limits on the nature of such a companion. Notably, by reconstructing images from sparse aperture masking data, using regularized maximum likelihood, \cite{2020ApJ...889L..24P} claimed that the resolved disk gap of HD 100546 was free of stellar companions. The contrast limits that were claimed in this analysis, however, were potentially overly optimistic, given that the reconstructed images were regularized to {explicitly} suppress (point-like) speckle features. More recently, \cite{2024A&A...682A.101S} used a point source companion model with sparse aperture masking data to explore the known disk gap for any potential point source companions{. This {model} assumes that only the central star and a point source companion significantly contribute to the observed measurement, ignoring any disk contribution. Thus,} despite recovering a high significance signal, the majority of what is seen is likely due to contamination from the {inner edge of the near side of the disk, along its minor axis. As noted by the authors, this is where disk emission is expected to be brightest in the near-infrared and therefore the recovered point source signal is likely to be due to the disk}. {Forward scattered light off a transition disk edge has previously been observed to produce signals that are similar to the signal produced by a point source in Sparse Aperture Masking Interferometry (SAM/AMI) data \citep[e.g.,][]{2013ApJ...762L..12C,2022ApJ...931....3B}.}

The bright disk emission present in these sparse aperture masking data sets of HD 100546 will obscure the signal of any moderate contrast point-like companion unless accounted for in a careful manner.
To mitigate this limitation, one approach is to fit geometrical models to the interferometric observables and perform Bayesian model comparison to determine the best representation of the data, {as was done for SAM observations of the disk sources LkCa 15} \citep{2022ApJ...931....3B,2023ApJ...953...55S} {and PDS 70} \citep{2024arXiv240413032B}. We apply this methodology to the archival VLT/SPHERE-IRDIS {SAM} data of HD 100546 that were previously published in \cite{2020ApJ...889L..24P} and \cite{2024A&A...682A.101S}. These data sets are ideal for this analysis due to the bright disk emission that well lends itself to geometrical modeling.
We aim to reveal the inner $\sim$10 au around HD 100546 A, inside the disk gap, to probe for companions that could produce at least some of the many observed disk features. 


This paper is structured as follows: \S\,\ref{sec:obs} outlines how the data were obtained and processed; \S\,\ref{sec:methods} details the model fitting procedure; \S\,\ref{sec:disc} outlines the results of applying these techniques and discusses the results, compares the various models that were explored and discusses the interpretation of the data. We close with a summary in \S\,\ref{sec:conc}.

\section{Observations} \label{sec:obs}

We analyze re-reduced VLT measurements of HD 100546 previously published by \cite{2020ApJ...889L..24P}, observed on the night of 15 May 2018 at {2.110 $\mu$m, $\Delta \lambda = 0.102$ $\mu$m, (K1) and 2.251 $\mu$m, $\Delta \lambda = 0.109$ $\mu$m, (K2)}  \citep[K12 dual-band imaging filters;][]{2010MNRAS.407...71V} as well as measurements previously published by \citet{2024A&A...682A.101S} and observed on 26 February and 17 March 2021 at {1.593 $\mu$m, $\Delta \lambda = 0.052$ $\mu$m,  (H2) and 1.667 $\mu$m, $\Delta \lambda = 0.054$ $\mu$m, (H3)} \citep[H23 dual-band imaging filters;][]{2010MNRAS.407...71V}. {The K12 data has a field of view (set by the longest and shortest baselines) of $\sim$35-260 mas (4-28 au) and the H23 data has a field of view of $\sim$25-195 mas (3-21 au).} All of these measurements were taken using the 7 hole (21 baseline) SPHERE-IRDIS \citep{2019A&A...631A.155B, 2008SPIE.7014E..3LD} SAM mode \citep{2011Msngr.146...18L,2010SPIE.7735E..1OT,2016SPIE.9907E..2TC}. Along with HD 100546, the following calibrator sources were observed using the same set-ups: HD 100901 (2018), HD 101966 (2018), HD 101531 (2018), HD 100560 (2021) and HD 101869 (2021). Due to the small separation in time between the 2021 observations, we treat these measurements as a single data set that we will henceforth refer to as the 2021 data.


{The observations were reduced following the basic steps outlined by \citet{2024A&A...682A.101S}, using \texttt{vlt-sphere} \citep{2020ascl.soft09002V}. }
The closure phases were extracted from the reduced data frames using \texttt{amical} \citep{2020SPIE11446E..11S}. Figure \ref{fig:all_data} shows the calibrated 
closure phases, which show a clear deviation from zero, which signifies a clear detection of relatively high contrast asymmetric emission around HD 100546. For the analysis presented in this work, only the closure phases are used due to systematics seen in the squared visibilities extracted from both data sets that can limit the probed companion contrasts \citep{2024A&A...682A.101S}.

\section{Methods} \label{sec:methods}

To represent the extended asymmetric disk emission in HD\,100546, we fit geometrical models directly to the interferometric observables. We follow the procedure outlined by \citet{2022ApJ...931....3B} for the transition disk LkCa 15, except where explicitly noted below. We employ a polar Gaussian ring, with peak amplitude fixed to the minor axis of the disk, to model the bright forward scattering component of the disk that is expected at near-infrared wavelengths \citep{2022arXiv220309991B}, and has previously been imaged around HD 100546 \citep[e.g.,][]{2016A&A...588A...8G}. As for LkCa 15, we adopt a polar Gaussian ring model given by
\begin{equation}
\label{eqn:pg}
    I_{PG}(r,\theta) = I_0 \exp\left(-\frac{(r-r_0)^2}{2\sigma_r^2} - \frac{(\theta-\theta_0)^2}{2\sigma_\theta^2}\right),
\end{equation}
where $r$ and $\theta$ are the radial and azimuthal coordinates, respectively.  $I_0$ is the peak amplitude of the ring, $r_0$ is the ring radius, $\theta_0$ specifies the position angle of the peak amplitude of the ring, and $\sigma_r$ and $\sigma_{\theta}$ specify the radial and azimuthal thickness, respectively.

Unlike with LkCa 15, we further test the inclusion of a second polar Gaussian disk component (also described by Equation \ref{eqn:pg}), with the same geometrical parameters as the forward scattering component, {but allowed to have an independent brightness, $I_1$, an independent azimuthal width, $\sigma_{\theta_1}$, as well as an independent azimuthal peak position, $\theta_1$, to model any additional scattered light from a disk asymmetry 
that is distinct from the forward scattering from the near-side (along the minor axis) of the disk edge.}

As was done by \cite{2022ApJ...931....3B}, we briefly explored the inclusion of an inner disk component allowed to have a small offset from the star (which we discuss more in Section \ref{sec:disc}) but found that any inner disk emission is likely unresolved. This is consistent with what has been seen with VLTI/GRAVITY at similar wavelengths, which found the semi-major axis of the inner disk of HD 100546 to be $\sim$2.6 mas \citep{2022AA...658A.183B}. So, for the purpose of this analysis, we assume that the inner disk is unresolved. 

Finally, we test the addition of a point source component, in addition to a single polar Gaussian ring (forward scattering) component. With this model, we test whether there is any evidence of compact asymmetric emission within the disk gap.

In the following, therefore, we compare the single polar Gaussian ring model (PG), the polar Gaussian ring plus a point source (PG+PS) and the two polar Gaussian rings model (2PG). 
We use dynamic nested sampling \citep{dyn2018} with \texttt{dynesty} \citep{2020MNRAS.493.3132S,sergey_koposov_2023_7600689} to sample from the model posterior, estimate model parameters and uncertainties, and calculate the {logarithm of the} Bayesian evidence (Log $\mathcal{Z}$). We use a Gaussian log-likelihood, using only the closure phase data, given by
\begin{equation}
    \label{like}
    \log{\mathcal{L}_{\phi}} = -\frac{1}{2} \sum_{i}^{n} \frac{(\phi_i - \hat{\phi}(V_i))^2}{\sigma_i^2} - \log{\sqrt{2\pi \sigma_i^2}},
\end{equation}
where $\phi_i$ are the measured closure phases, $\sigma_i$ are their associated uncertainties, and $\hat{\phi}(V_i)$ are the model closure phases. 
For all of the fits, we use uniform priors on all of the model parameters, except for the position angle of the disk, which uses a Gaussian prior with a mean of 323$^{\circ}$ and a standard deviation of 5$^{\circ}$, reflecting a loose estimate of the distribution of literature measurements of the disk orientation \citep[][]{2019ApJ...871...48P,2022ApJ...933L...4C,2022AA...658A.183B}.

{For all three models, we perform a joint analysis by fixing the geometry of the main forward scattering disk component between epochs while allowing the brightness of the disk to vary between each of the filters.} For the point source component, we fit for a single position at each of the two epochs (2018 and 2021), allowing for potential companion motion, and fit for individual companion-to-star contrasts at each of the four filters. {For the second ring component of the 2PG model, we allow the azimuthal location ($\theta_1$), extent ($\sigma_{\theta,1}$) and brightness ($I_1$) of the second polar Gaussian ring component to vary between the two epochs.}


\section{Results \& Discussion} \label{sec:disc}

\begin{figure*}
\centering
\includegraphics[width=1\linewidth]{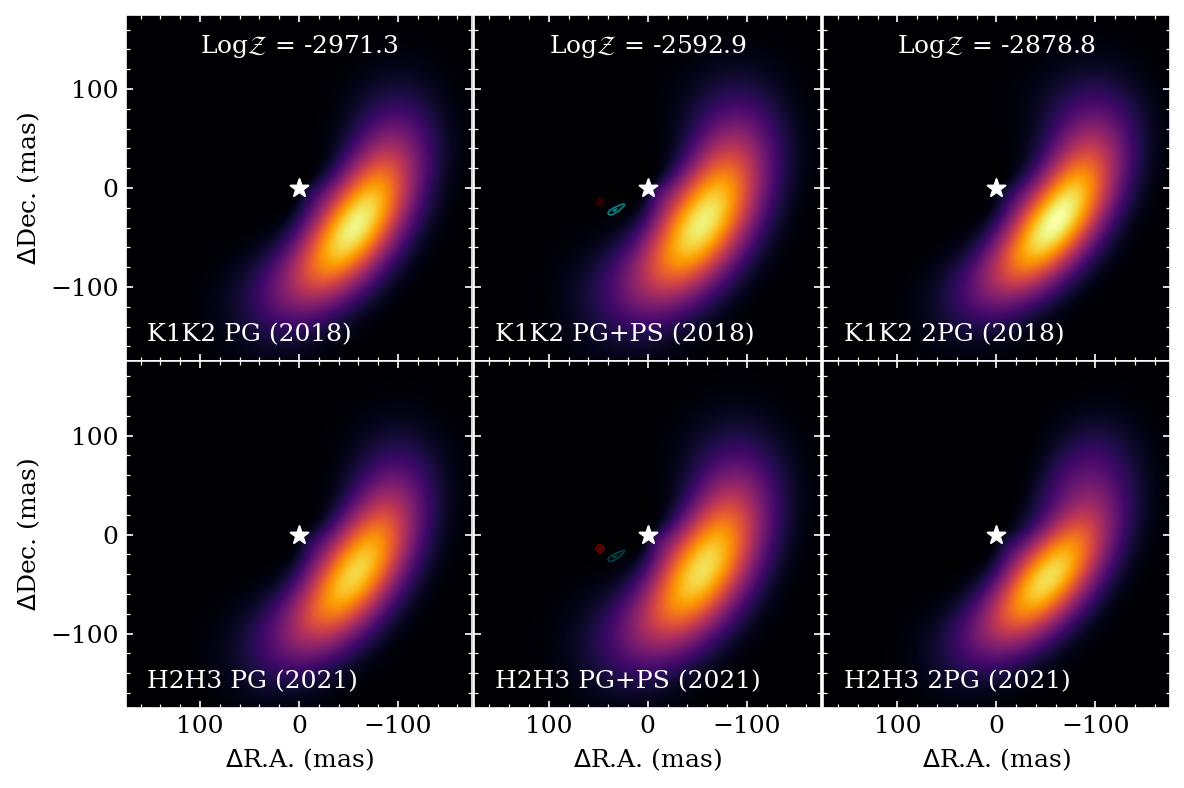}
\caption{{Median geometrical models from the PG (left), PG+PS (middle) and 2PG (right) joint fits to the closure phase data, calculated using the median parameters displayed in Table \ref{tab:comp_fit_par}. The top row shows the 2018 results and the bottom row shows the 2021 results. For the PG+PS model, the 1 and 3 $\sigma$ credible intervals on the companion position are shown by the teal (2018) and maroon (2021) contours. The semi-transparent contours denote the credible intervals of the other epoch, to illustrate the significance of the motion between the two epochs. The Bayesian evidence (Log$\mathcal{Z}$), calculated with nested sampling, are shown for each model at the top of each row. With a Log$\mathcal{Z}$ = -2592.9, 378.4 greater than the PG model, and 285.9 greater than the 2PG model, the PG+PS is resoundingly the best representation of the data of the explored models, supporting the interpretation that there is asymmetric emission originating in the disk gap.}}
\label{fig:jointfits_plot}
\end{figure*}

\subsection{Results}




\begin{deluxetable}{ccc}
    \label{tab:model_pars}
    \tablecaption{Model parameters. }
    
    \tablewidth{0pt}
    
    \tablehead{&Prior&Credible Range}
    \startdata
    \hline
    {PG}&&\\
    \hline
     {log $I_{0,H2}$ (arb.)}&$\mathcal{U}$(-6, -2)& $-4.505^{+0.008}_{-0.008}$\\
     {log $I_{0,H3}$ (arb.)}&$\mathcal{U}$(-6, -2)& $-4.509^{+0.006}_{-0.006}$\\
     {log $I_{0,K1}$ (arb.)}&$\mathcal{U}$(-6, -2)& $-4.474^{+0.005}_{-0.005}$\\
     {log $I_{0,K2}$ (arb.)}&$\mathcal{U}$(-6, -2)& $-4.449^{+0.005}_{-0.005}$\\
     {$r_0$ (mas)}&$\mathcal{U}$(0.05,0.15)& $0.102^{+0.001}_{-0.001}$\\
     {$i$ ($^{\circ}$)}&$\mathcal{U}$(30,60)& $47.7^{+0.5}_{-0.5}$\\
     {P.A. ($^{\circ}$)}&$\mathcal{N}(323, 5)$& $325.6^{+0.2}_{-0.2}$\\
     {FWHM$_r$ (mas)}&$\mathcal{U}$(0.03, 0.2)& $0.101^{+0.001}_{-0.001}$\\
     {FWHM$_{\theta}$ ($^{\circ}$)}&$\mathcal{U}$(40, 180)& $92.9^{+1.2}_{-1.2}$\\
     \hline
    {PG+PS}$^*$&&\\
    \hline
    {log $I_{0,H2}$ (arb.)}&$\mathcal{U}$(-6, -2)& $-4.495^{+0.008}_{-0.009}$\\
     {log $I_{0,H3}$ (arb.)}&$\mathcal{U}$(-6, -2)& $-4.500^{+0.007}_{-0.007}$\\
     {log $I_{0,K1}$ (arb.)}&$\mathcal{U}$(-6, -2)& $-4.478^{+0.005}_{-0.005}$\\
     {log $I_{0,K2}$ (arb.)}&$\mathcal{U}$(-6, -2)& $-4.453^{+0.005}_{-0.005}$\\
     {$r_0$ (mas)}&$\mathcal{U}$(0.05,0.15)& $0.093^{+0.001}_{-0.001}$\\
     {$i$ ($^{\circ}$)}&$\mathcal{U}$(30,60)& $45.1^{+0.6}_{-0.6}$\\
     {P.A. ($^{\circ}$)}&$\mathcal{N}(323, 5)$& $327.9^{+0.3}_{-0.3}$\\
     {FWHM$_r$ (mas)}&$\mathcal{U}$(0.03, 0.2)& $0.105^{+0.001}_{-0.001}$\\
     {FWHM$_{\theta}$ ($^{\circ}$)}&$\mathcal{U}$(40, 180)& $101.6^{+1.8}_{-1.6}$\\
    \hline
    {2PG}&&\\
    \hline
    {log $I_{0,H2}$ (arb.)}&$\mathcal{U}$(-6, -2)& $-4.497^{+0.008}_{-0.008}$\\
     {log $I_{0,H3}$ (arb.)}&$\mathcal{U}$(-6, -2)& $-4.499^{+0.007}_{-0.007}$\\
     {log $I_{0,K1}$ (arb.)}&$\mathcal{U}$(-6, -2)& $-4.58^{+0.04}_{-0.08}$\\
     {log $I_{0,K2}$ (arb.)}&$\mathcal{U}$(-6, -2)& $-4.53^{+0.03}_{-0.06}$\\
     {$r_0$ (mas)}&$\mathcal{U}$(0.05,0.15)& $0.104^{+0.001}_{-0.001}$\\
     {$i$ ($^{\circ}$)}&$\mathcal{U}$(30,60)& $48.9^{+0.5}_{-0.5}$\\
     {P.A. ($^{\circ}$)}&$\mathcal{N}(323, 5)$& $320.0^{+0.5}_{-0.5}$\\
     {FWHM$_r$ (mas)}&$\mathcal{U}$(0.03, 0.2)& $0.099^{+0.001}_{-0.001}$\\
     {FWHM$_{\theta}$ ($^{\circ}$)}&$\mathcal{U}$(40, 180)& $83.2^{+1.7}_{-1.8}$\\
    {log $I_{1,H2}$ (arb.)}&$\mathcal{U}$(-7, -3)& $-5.23^{+0.05}_{-0.05}$\\
     {log $I_{1,H3}$ (arb.)}&$\mathcal{U}$(-7, -3)& $-5.34^{+0.08}_{-0.09}$\\
     {log $I_{1,K1}$ (arb.)}&$\mathcal{U}$(-7, -3)& $-4.95^{+0.11}_{-0.07}$\\
     {log $I_{1,K2}$ (arb.)}&$\mathcal{U}$(-7, -3)& $-5.01^{+0.11}_{-0.07}$\\
     {$\theta_{1,2021}$ ($^{\circ}$)}&$\mathcal{U}$(-180, 180)& $73^{+2}_{-2}$\\
     {$\theta_{1,2018}$ ($^{\circ}$)}&$\mathcal{U}$(-180, 180)& $35^{+9}_{-10}$\\
     {FWHM$_{\theta,1,2021}$ ($^{\circ}$)}&$\mathcal{U}$(40, 180)& $48^{+7}_{-5}$\\
     {FWHM$_{\theta,1,2018}$ ($^{\circ}$)}&$\mathcal{U}$(40, 180)& $96^{+7}_{-9}$\\
     \hline
    \enddata
    \footnotesize{Notes: Credible range shows the 50$^{\mathrm{th}}$ percentile plus the 84$^{\mathrm{th}}$ and minus 16$^{\mathrm{th}}$ percentiles. The reported FWHM values are equal to $ \sigma \sqrt{8\ln{2}}$.
    
    $^*$ The companion parameters of the PG+PS model are shown in Table \ref{tab:comp_fit_par}.}
\end{deluxetable}

\begin{deluxetable}{cccc}
    \label{tab:comp_fit_par}
    \tablecaption{Companion parameters from the PG+PS model}
    
    \tablewidth{0pt}
    
    \tablehead{Band/Epoch&log$_{10}$Contrast&Separation & Position Angle\\
    \nocolhead{name} &\nocolhead{name} &\colhead{(mas)}& \colhead{($^{\circ}$)}}
    \startdata
     \hline
      {K1 (2018)}&$-2.56^{+0.06}_{-0.04}$& $39.9^{+2.8}_{-3.3}$ & $124.1^{+1.0}_{-1.0}$\\
     {K2 (2018)}&$-2.57^{+0.06}_{-0.05}$& $-$ & $-$ \\
     {H2 (2021)}&$-2.88^{+0.04}_{-0.04}$& $50.0^{+1.0}_{-1.1}$ & $106.4^{+1.4}_{-1.4}$\\
     {H3 (2021)}&$-2.90^{+0.04}_{-0.04}$& $-$ & $-$ \\
     \hline
    \enddata
    \footnotesize{Note: Fit values denote the 50$^{\mathrm{th}}$ percentile plus the 84$^{\mathrm{th}}$ and minus 16$^{\mathrm{th}}$ percentiles. {Uniform priors were used on all of the companion parameters, given by: log$_{10}$Contrast\,$\sim \mathcal{U}(-4,-1)$, Separation\,$\sim \mathcal{U}(0,200)$ and Position Angle\,$\sim \mathcal{U}(0,360)$. For the contrast prior, we exclude very low contrast values inconsistent with non-detections in previous VLTI observations \citep[e.g.,][]{2017AA...599A..85L,2022AA...658A.183B}}.}
\end{deluxetable}


{Figure \ref{fig:jointfits_plot} shows the geometrical models, using the median model parameters from nested sampling, by type (rows), filter/epoch (columns). All three models show a similar general ring structure. Notably, there are moderate differences in the position angle of the disk between models (Table \ref{tab:model_pars}). The radius of the PG+PS model is smaller than for the PG and 2PG models, but this is compensated for by the PG+PS model having a lower inclination. Table \ref{tab:comp_fit_par} shows the companion parameters extracted from the PG+PS model. For this model, it is notable that a statistically significant clockwise offset of $\sim$18 degrees in position angle, from a position angle of $124.1^{+1.0}_{-1.0}$ degrees to $106.4^{+1.4}_{-1.4}$ degrees and $\sim$10 mas in separation, from a separation of $39.9^{+2.8}_{-3.3}$ mas (projected separation of $\sim$4.3 au) to $50.0^{+1.0}_{-1.1}$ mas (projected separation of $\sim$5.4 au)  is seen between the 2018 and 2021 epochs. For the 2PG model, the location of the peak of the secondary disk asymmetry moves counter-clockwise by $\sim 40$ degrees from an azimuthal location of $35^{+9}_{-10}$ degrees to $73^{+2}_{-2}$ degrees, relative to the near-side (forward scattering peak) minor axis of the disk model. It is worth noting that the location of this model disk asymmetry is on the opposite side (relative to the near-side minor axis of the disk) as the spiral feature, originating along the disk edge major axis to the south-east, previously observed in the near infrared \citep{2017AJ....153..264F,2022ApJ...936L...4N}. }

\subsection{Analysis of the disk only model residuals}

{To supplement the point source plus polar Gaussian (PG+PS) model fit results, we additionally analyze the residuals from the single polar Gaussian (PG) fits. We visualize the remaining asymmetry left in the data by computing a reduced chi-squared, $\chi^2_{r}$, map for a point source (binary) model fit to these residuals. The $\chi^2_{r}$ was evaluated by adding the complex {visibilities} of the PG model, calculated from the median parameters listed in Table \ref{tab:model_pars} for the PG model fit, to the complex {visibilities} of the fitted binary model, then determining the closure phases and calculating the likelihood. The map was produced by evaluating the $\chi^2_{r}$ for a grid of 250 points in contrast, separated logarithmically from 10$^{-6}$ to 10$^{-1}$, and 1 mas cells in right ascension and declination, from -125 to 125 mas, then plotting a map of the minimum $\chi^2_{r}$ for each spatial grid point. }

\begin{figure*}
\centering
\includegraphics[width=1\linewidth]{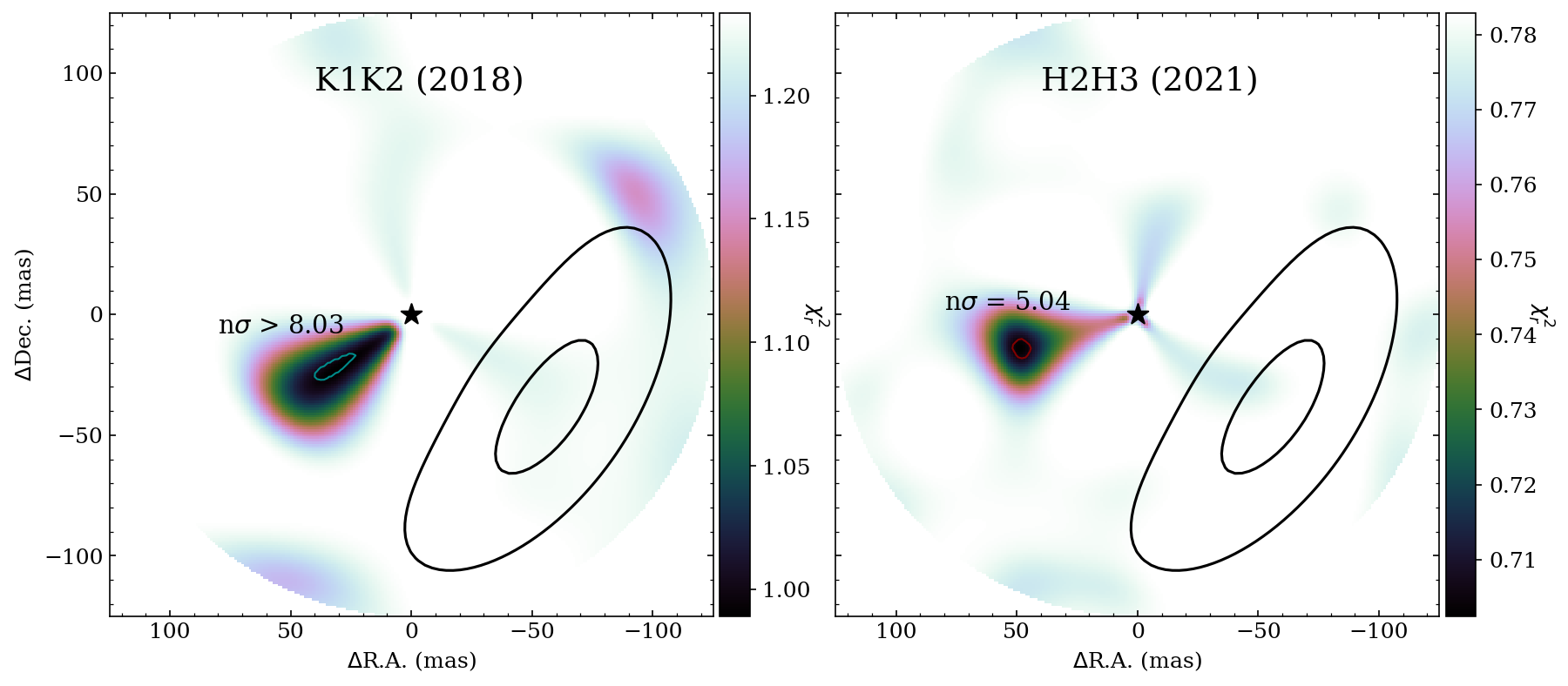}
\caption{{Reduced chi-squared ($\chi^2_{r}$) maps, calculated by fitting a point source (binary) model to the residuals of the PG model fit, at each location (with a step size of 1 mas) within a radius of 125 mas. The significance of the 2018 detection is a $>$8$\sigma$ result and the 2021 detection is a 5$\sigma$ result, using the method by \cite{2015A&A...579A..68G}. The 3$\sigma$ credible interval contours on the location of the point source at each epoch are denoted by the teal (2018, left) and maroon (2021, right) contours.} The 90\% and 50\% contours of the disk from the single polar Gaussian (PG) fit are denoted by the black lines.}
\label{fig:chi2_plot}
\end{figure*}

{The $\chi^2_{r}$ maps are shown in Figure \ref{fig:chi2_plot}. 
The global minima are seen at similar positions and contrasts to what was found with the PG+PS model. In the 2018 maps, the expected degeneracy between contrast and separation is seen, showing a $\chi^2_{r}$ plateau (also seen in the bottom right of the PG+PS corner plot shown in Figure \ref{fig:corner_plot}, in Appendix \ref{sec:post}), along the same position angle. Using Equation 4 from \cite{2024A&A...682A.101S}  \citep[see also][]{2015A&A...579A..68G}, comparing the $\chi^2_{r}$ of the binary model to that of a star only model, we calculate the significance of the $\chi^2_{r}$ minima.
For the 2018 data, we find that 
the binary model is preferred with $>8.03 \sigma$ significance (this is the numerical upper limit of this method), and the binary model is preferred at the 5.04$\sigma$ level for the 2021 data.}

\subsection{Model Comparison and Interpretation}

\begin{figure*}
\centering
\includegraphics[width=1\linewidth]{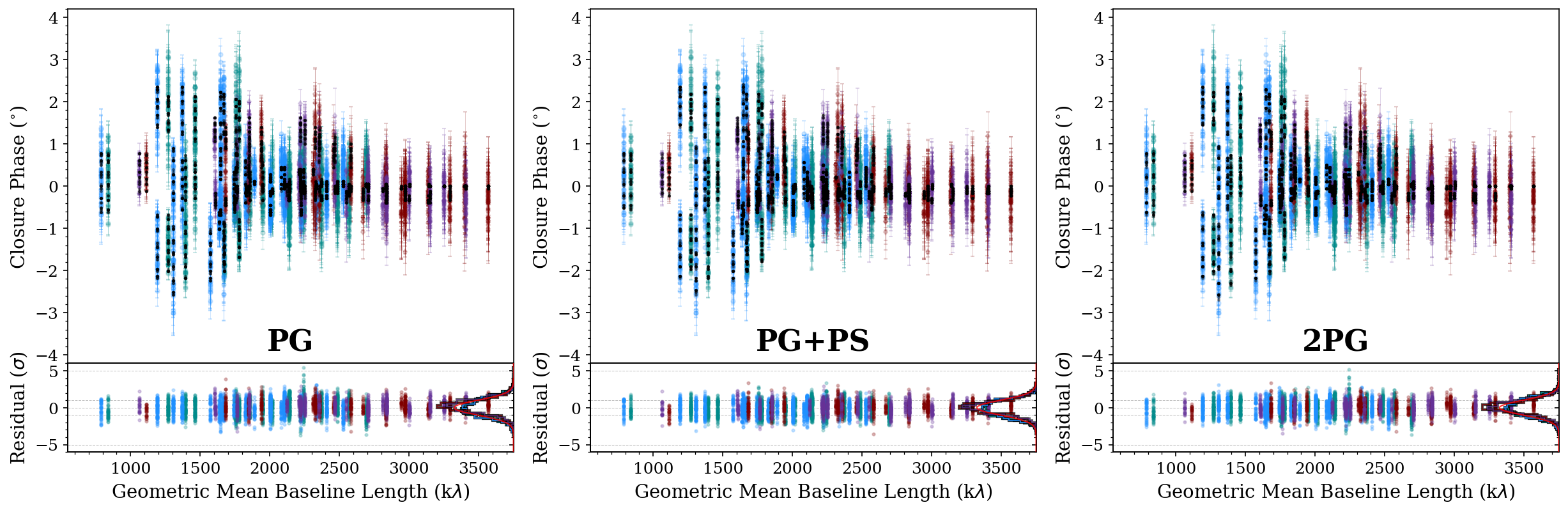}
\caption{{Closure phase data (using the same colours as in Figure \ref{fig:all_data}) and model (black), for the PG (left), PG+PS (middle) and 2PG (right) models. The bottom panels in each plot show the noise normalized residuals, with the binned residuals shown on the right and the red curve denoting a unit normal distribution.}}
\label{fig:corr_plot}
\end{figure*}

{The PG, PG+PS and 2PG models all fit the data reasonably well, as can be seen in Figure \ref{fig:corr_plot} which compares the closure phases calculated from each of the models to the data. From the nested sampling results, we find a maximized likelihood,  $\log{\mathcal{L}_{\phi}}^{*}$  of -2929.3 for the PG model, -2519.8 for the PG+PS model and -2811.2 for the 2PG. It is not unexpected that the PG+PS and 2PG models have higher likelihoods than the PG model because they are both supersets of the PG model and more expressive. However, it is notable that the PG+PS model fits the data significantly better than the 2PG model, as they both have 17 free parameters. We can also compare the three models using Bayes factors (the ratio of Bayesian model evidences), which includes a penalty for overly complex models \citep{doi:10.1080/01621459.1995.10476572,bayesevid2011}. We find Bayesian evidence values, Log$\mathcal{Z}$, of -2971.3, -2592.9 and -2878.8 for the PG, PG+PS and 2PG models, respectively. We find a log-Bayes factor of 378.4 comparing the PG+PS model to the PG model, and 92.5 comparing the 2PG model to the PG model. Thus, the polar Gaussian ring plus point source companion model (PG+PS) is the best representation of the data, out of the three models.}
Due to the complex morphology seen in scattered light, however, this should not be taken as unambiguous evidence of a sub-stellar companion in the system. While compelling, the results must be considered with care, due to the limited and simple geometrical models that are tested. 

Recognizing our caution against making a decisive claim for the nature of the disk gap around HD 100546, we are able to rule out some of the parameter and model spaces. The results from the 2PG model, combined with the consistent, localized nature of the point source fits, well inside the disk gap, all but rule out the likelihood that the observed closure phase asymmetry is due to emission from the outer disk. From the PG+PS fits, and our analysis of the residuals of the PG model residuals (Figure \ref{fig:chi2_plot}), a point source companion is a plausible explanation for the observed residual asymmetric signal. We note however that it is also possible that the signal could be being produced by an extended asymmetric structure associated with the inner disk but extending out to the locations derived with the point source model. However, this scenario is unlikely since our measured point source separations ($\sim$40-50 mas, Table \ref{tab:comp_fit_par}) from the central source are larger than most estimates of the inner disk size \citep[e.g.,][]{2014A&A...562A.101P,2018ApJ...865..137J,2022AA...658A.183B}.

The only scenario that we are not able to significantly constrain is that of a disk asymmetry, at small ($<$1 au, $<$9 mas) separation from the central star, that is bright enough to have a non-negligible contribution to the measured closure phases (this is possible due to the contrast/separation degeneracy, which, as has previously noted, can be clearly seen in {the bottom right of Figure \ref{fig:corner_plot}, in Appendix \ref{sec:post}, as well as in the left panel of Figure \ref{fig:chi2_plot}. Such degeneracy} occurs when working with closure phases within $\sim \lambda / B_{\text{max}}$, where $\lambda$ is the observing wavelength and $B_{\text{max}}$ is the longest baseline). This scenario is extremely difficult to test using the SAM data that we have analyzed, due to the inner working angle of the SAM data being $\sim$20-25 mas (0.5$\lambda / B_{\text{max}}$), and the need for a complex inner disk model or an unbiased non-parametric image reconstruction, which is itself a challenge due to the many degeneracies inherent to the ill-posed, non-linear problem of reconstructing images non-parametrically using only closure phases. Despite these limitations, we also explored an inner disk model, parameterized by an elliptical Gaussian that was allowed to be offset from the star, as was contemplated by \cite{2022ApJ...931....3B} for LkCa 15.
However, unlike the LkCa 15 result, we find that this component does not remain bound to the star when allowed to have an $x, y$ offset. This disfavours the hypothesis of the observed signal being due to an extremely bright, small inner disk asymmetry, though we cannot rule it out entirely using these tests.

\begin{figure*}
\centering
\includegraphics[width=\linewidth]{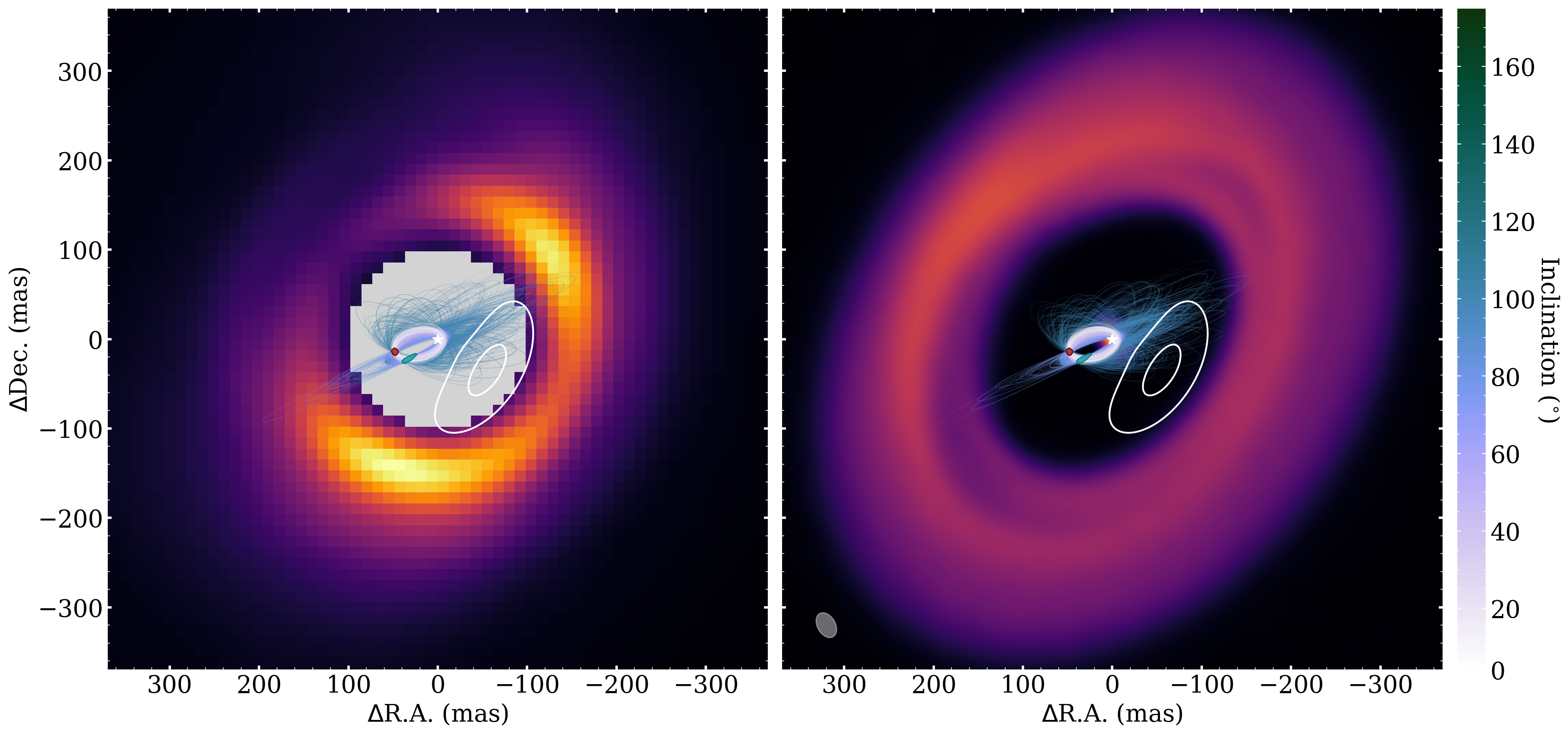}
\caption{The white contours show the {median model} disk component of the PG+PS model plotted on top of a \textit{K$_s$}-band total intensity image of HD 100546 from \cite{2023A&A...680A.114R} (left) and ALMA band 6 data of HD 100546 from \cite{2022ApJ...936L...4N} (right). The images are shown on a linear scale. In both panels, the location of the central star is denoted by the white star and the 3 $\sigma$ credible interval on the point source positions from the joint fit to all datasets with the PG+PS model are shown by the teal (2018) and maroon (2021) contours. The beamsize of the ALMA image is shown in the bottom left corner of the right panel. 1000 randomly drawn orbits from the orbital fit posteriors are plotted on top of both images. Note that in the right panel the inner disk in the ALMA image is almost entirely occulted by the plotted orbits.}
\label{fig:joint_fit}
\end{figure*}

\subsection{Derived Disk and Companion Features}

In the {left} panel of Figure \ref{fig:joint_fit}, we show the contours of the PG+PS model, plotted on top of a \textit{K$_s$}-band, coronographic, total intensity image of HD 100546, from \cite{2023A&A...680A.114R}. The orientation of our model closely matches what is seen in the coronographic image, however, the gap edge (along the minor axis of the disk) appears to be hidden behind the coronograph. This explains why the peak of the emission in the coronographic image lies close to the major axis, as the emission may be from the illuminated disk wall, whereas the emission from the minor axis, on the near side of the disk (to the south-west), is likely from the disk surface -- so we are seeing a trade off between scattering angle and seeing the disk wall and/or disk surface. The location of the peak of the emission found with our geometrical models (which represents the top of the inner edge of the near-side of the disk) supports this hypothesis, as it lies well within the region hidden behind the coronograph in the image from \cite{2023A&A...680A.114R}. Additionally, the location of the emission maximum found from our modeling closely matches the location of the gap edge seen in the reconstructed image of the 2018 VLT/SPHERE-IRDIS SAM dataset  \citep{2020ApJ...889L..24P}. {For completeness, in the right panel of Figure \ref{fig:joint_fit}, we show the contours of the PG+PS model plotted on top of an ALMA band 6 image of HD 100546 from \cite{2022ApJ...936L...4N}}.

To estimate the nature of the companion candidate emission that we detect, we first convert our measured contrasts to mass estimates using the \texttt{atmo-ceq} atmosphere and evolution models for cool T-Y brown dwarfs and giant exoplanets \citep{2020A&A...637A..38P} with \texttt{species} \citep{2020A&A...635A.182S}, then explore the orbital parameter space consistent with our measured astrometry using \texttt{Octofitter} \citep{2023AJ....166..164T}. 

As was previously discussed, the measured point-source contrasts suffer from the well known closure phase contrast/separation degeneracy (especially in \textit{K}-band). Another consequence of this degeneracy coupled with the fact that the disk signal and point source signal are not entirely orthogonal (Figure \ref{fig:corner_plot}, Appendix \ref{sec:post}) is that it is impossible to entirely disentangle the disk signal from any companion signal due to the simplistic disk model employed in this work. 

Acknowledging the chance that our measured contrasts are somewhat biased, we estimate the mass of the observed object, using our measured contrasts from the jointly fit model, assuming the object has an age of 5 Myr \citep{2019AJ....157..159A}, and is not locally extincted due to the object being embedded. To convert the measured contrasts to fluxes, we use 2MASS \citep{2006AJ....131.1163S} \textit{H} and \textit{K$_s$}-band magnitudes. We expect that the error in using these estimates (as opposed to fitting a stellar SED model) for the stellar flux is negligible compared to the systematics in our reported contrasts, so we solely report the range of masses we find using this procedure. From this approach, we estimate a mass range of $\sim$25-50 M$_{\text{J}}$. The lower bound is from the H3-band measurement and the upper bound is from the K1-band measurement. To measure a reliable spectrum of the companion candidate, follow-up observations with VLTI/GRAVITY dual field mode \citep[e.g.,][]{2024A&A...686A.258P} will be necessary. This will require first a more precise orbit determined from additional follow-up epochs of SAM data. 

{We explore the orbital parameter space of a companion to HD 100546 that is consistent with the measured astrometry of the point source from the PG+PS model (Table \ref{tab:comp_fit_par}) with \texttt{Octofitter} \citep{2023AJ....166..164T}, using Non-Reversible Parallel Tempering \citep{https://doi.org/10.1111/rssb.12464,surjanovic2023pigeons}. The priors we use are listed in Table \ref{tab:orbit_priors}. For the stellar mass, we use a conservatively wide Gaussian prior, due to a discrepancy in stellar mass estimates of HD 100546 A derived with different methods. The extracted orbital parameters  are also listed in Table \ref{tab:orbit_priors}. 
In Figure \ref{fig:joint_fit}, we show randomly drawn orbits from the orbital posterior, with contours denoting the disk component of the PG+PS model, plotted on top of \textit{K$_s$}-band, coronographic, total intensity image of HD 100546, from \cite{2023A&A...680A.114R} (left) and ALMA band 6 data of HD 100546 from \cite{2022ApJ...936L...4N}.}

\begin{figure}
\centering
\includegraphics[width=1\linewidth]{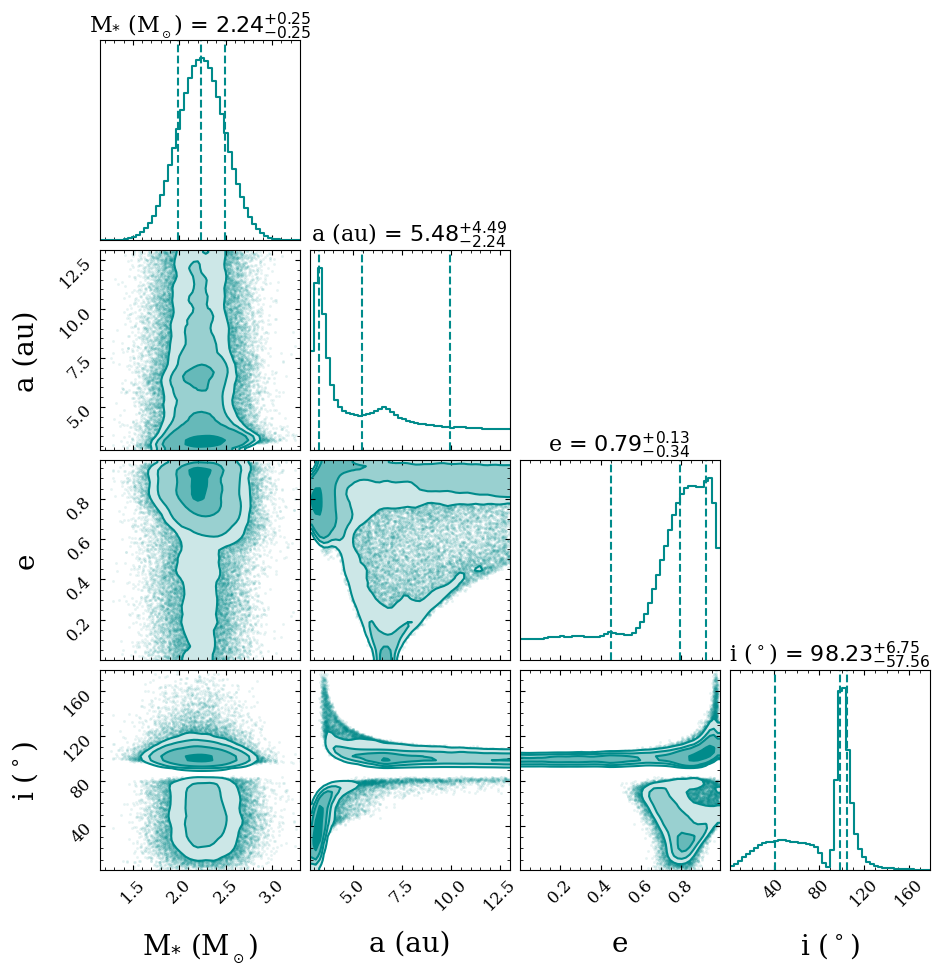}
\caption{{Orbital analysis corner plot, showing only the stellar mass, semi-major axis, eccentricity and inclination distributions.}}
\label{fig:orbit_corner_plot}
\end{figure}

{We find orbital solutions with inclination 98$^{+7}_{-58}$, and eccentricity $0.8^{+0.1}_{-0.3}$. We find a multi-modal distribution, both with a broad peak at $i \sim$50 degrees and another at $i \sim$100 degrees. This bimodality can be clearly seen in the corner plot shown in Figure \ref{fig:orbit_corner_plot}. The lower inclination orbits largely have smaller semi-major axes and higher eccentricities. A portion of the lower inclination orbits also correspond to counter-clockwise orbits (in the plane of the sky), implying that our measured astrometry is consistent with a companion near its apastron on subsequent orbits around HD 100546 A. For the higher inclination mode, the orbits are counterclockwise, with the majority of orbits being consistent with a companion that has not completed a full orbit around HD 100546 A between detections.}

\begin{deluxetable}{ccc}
    \label{tab:orbit_priors}
    \tablecaption{Point source orbital parameters. }
    
    \tablewidth{0pt}
    
    \tablehead{Orbital Element&Prior&Credible Range}
    \startdata
     {$a$ (au)}&$\mathcal{U}$(1, 13)& $5^{+4}_{-2}$\\
     {$e$}&$\mathcal{U}$(0, 0.99)& $0.8^{+0.1}_{-0.3}$\\
     {$i$ ($^{\circ}$)}&Sine()& $98^{+7}_{-58}$\\
     {$\omega$ ($^{\circ}$)}&$\mathcal{U}$(0, 360)& $179^{+109}_{-113}$\\
     {$\Omega$ ($^{\circ}$)}&$\mathcal{U}$(0, 360)& $198^{+97}_{-105}$\\
     {$\theta^1$}&$\mathcal{U}$(0, 360)& $124.2^{+1.0}_{-1.0}$\\
     {M$_s$ (M$_{\odot}$)}&$\mathcal{N}$(2.25, 0.25)& $2.24^{+0.25}_{-0.25}$\\
     {Parallax (mas)}&$\mathcal{N}$(9.25, 0.04)$^{2}$& $9.25^{+0.04}_{-0.04}$\\
     \hline
    \enddata
    \footnotesize{Notes: Fit values denote the 50$^{\mathrm{th}}$ percentile plus the 84$^{\mathrm{th}}$ and minus 16$^{\mathrm{th}}$ percentiles. We use the mean, standard deviation and correlation calculated from the joint fit posterior described in Table \ref{tab:comp_fit_par} and shown in Figure \ref{fig:corner_plot}, in Appendix \ref{sec:post}. For the 2018 data, these values are sep. $=$ 39.6 $\pm$ 3.1 mas, P.A. $=$ 124.1 $\pm$ 1.0 degrees, with a correlation of -0.296. For the 2021 data, they are sep. $=$ 50.0 $\pm$ 1.1 mas, P.A. $=$ 106.4 $\pm$ 1.4 degrees, with a correlation of -0.022.
    
    $^1$ The position angle of the companion at a reference epoch of mjd $=$ 58254.
    
    $^2$ Reference: \cite{2023AA...674A...1G}.}
\end{deluxetable}

\section{Conclusions} \label{sec:conc}

In this work, we re-analyzed two epochs of archival VLT/SPHERE-IRDIS SAM data of HD 100546 observed in 2018 and 2021. We fit multiple distinct geometrical models to the data and found that the polar Gaussian plus point source model (PG+PS), representing forward scattering from the disk edge and an unresolved compact source of emission in the disk gap, was the best representation of the data. We found that this tentatively detected point source moved by $\sim$10 mas in separation and $\sim$18 degrees in position angle, over the three years between the two observations. From the measured contrasts, we estimated the mass of this object to be $\sim$25-50 M$_{\text{J}}$. 
Our measured astrometry is consistent with either an object orbiting with a high eccentricity ($e \gtrapprox 0.65$), orbiting close to the disk plane, or an object with any eccentricity, orbiting at a large inclination relative to the disk ($\Delta i \sim 60$ degrees). 
However, at both epochs, the location of the point source is near the diffraction limit of the SAM data (especially in \textit{K}-band), where there is a degeneracy between separation and contrast, which means that the observed signal could possibly be reproduced by a bright asymmetry associated with the inner disk, which we cannot rule out with the analyzed data. Since we {cannot} definitively rule out this scenario, to robustly confirm this candidate, follow-up high-contrast observations will be necessary to fully understand the observed signal and distinguish between a low-mass (sub-)stellar companion versus a complex disk {feature}. These two potential scenarios can likely be distinguished by either re-observing HD 100546 with VLT/SPHERE SAM to confirm orbital motion, or observing HD 100546 with VLTI/GRAVITY (combining new observations with archival data) to constrain the size and morphology of the inner disk as a function of time. 

\begin{acknowledgments}
D.B. and D.J. acknowledge the support of the Natural
Sciences and Engineering Research Council of Canada
(NSERC). D.J. also acknowledges support from NRC Canada. This project has received funding from the European Research Council (ERC) under the European Union's Horizon 2020 research and innovation programme (PROTOPLANETS, grant agreement No.~101002188).
\end{acknowledgments}

%

\vspace{5mm}


\software{\texttt{JAX} \citep{jax2018github}, \texttt{dynesty} \citep{2020MNRAS.493.3132S,sergey_koposov_2023_7600689}, \texttt{astropy} \citep{2013A&A...558A..33A,2018AJ....156..123A}, \texttt{matplotlib} \citep{Hunter:2007}, \texttt{corner} \citep{corner}, \texttt{species} \citep{2020A&A...635A.182S}, \texttt{NumPy} \citep{harris2020array}.}



\appendix

\section{PG+PS Posterior Distribution}
\label{sec:post}

\begin{figure*}
\centering
\includegraphics[width=1\linewidth]{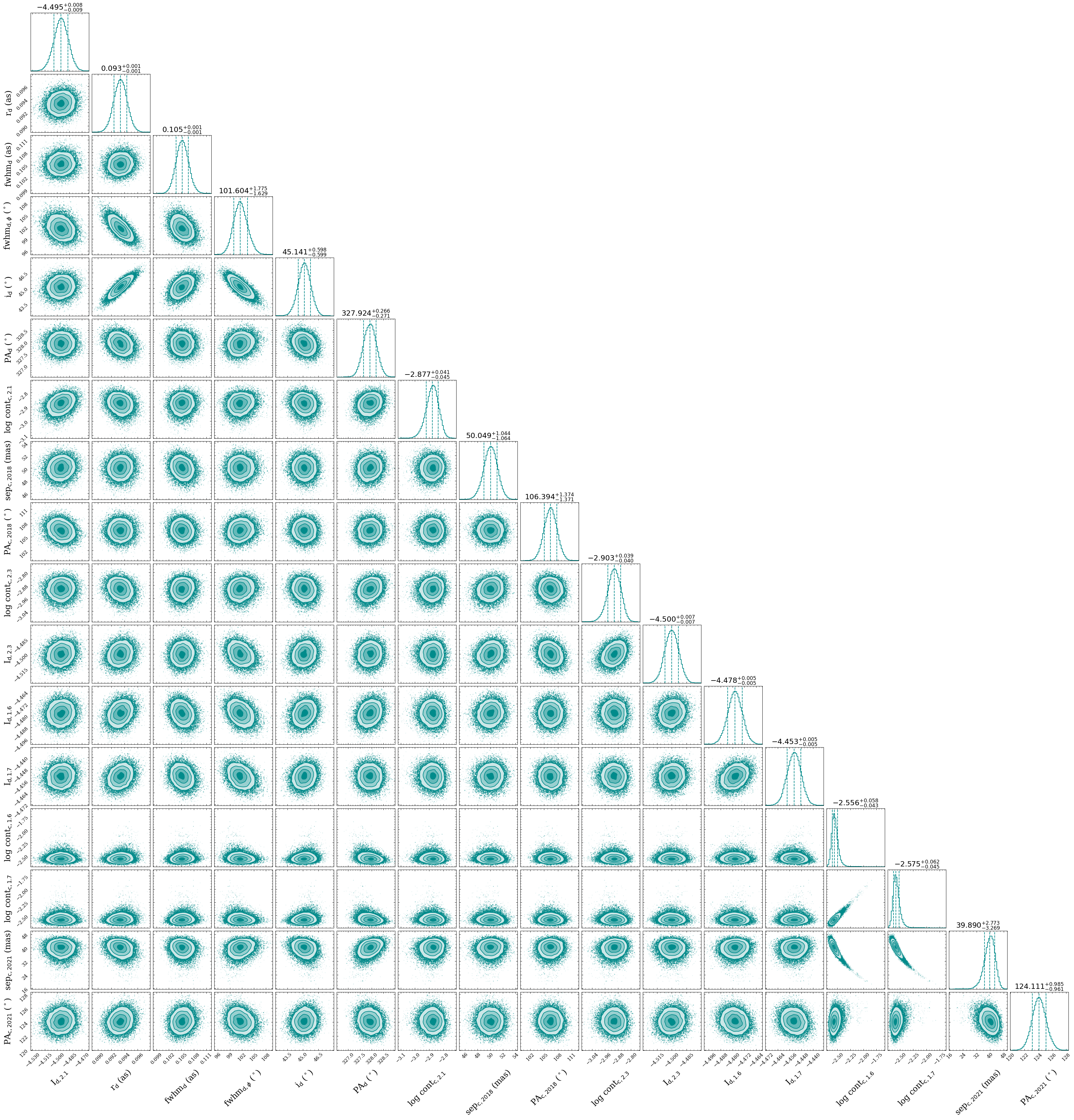}
\caption{Corner plot from the joint fit of the PG+PS model, with the disk geometry fixed between epochs, and the companion location allowed to vary.}
\label{fig:corner_plot}
\end{figure*}

Figure \ref{fig:corner_plot} shows the posterior distribution from the joint point source plus polar Gaussian (PG+PS) fit to all of the data.

\bibliography{sample631}{}

\begin{thebibliography}{}
\expandafter\ifx\csname natexlab\endcsname\relax\def\natexlab#1{#1}\fi
\providecommand{\url}[1]{\href{#1}{#1}}
\providecommand{\dodoi}[1]{doi:~\href{http://doi.org/#1}{\nolinkurl{#1}}}
\providecommand{\doeprint}[1]{\href{http://ascl.net/#1}{\nolinkurl{http://ascl.net/#1}}}
\providecommand{\doarXiv}[1]{\href{https://arxiv.org/abs/#1}{\nolinkurl{https://arxiv.org/abs/#1}}}

\bibitem[{{Arun} {et~al.}(2019){Arun}, {Mathew}, {Manoj}, {Ujjwal}, {Kartha}, {Viswanath}, {Narang}, \& {Paul}}]{2019AJ....157..159A}
{Arun}, R., {Mathew}, B., {Manoj}, P., {et~al.} 2019, \aj, 157, 159, \dodoi{10.3847/1538-3881/ab0ca1}

\bibitem[{{Astropy Collaboration} {et~al.}(2013){Astropy Collaboration}, {Robitaille}, {Tollerud}, {Greenfield}, {Droettboom}, {Bray}, {Aldcroft}, {Davis}, {Ginsburg}, {Price-Whelan}, {Kerzendorf}, {Conley}, {Crighton}, {Barbary}, {Muna}, {Ferguson}, {Grollier}, {Parikh}, {Nair}, {Unther}, {Deil}, {Woillez}, {Conseil}, {Kramer}, {Turner}, {Singer}, {Fox}, {Weaver}, {Zabalza}, {Edwards}, {Azalee Bostroem}, {Burke}, {Casey}, {Crawford}, {Dencheva}, {Ely}, {Jenness}, {Labrie}, {Lim}, {Pierfederici}, {Pontzen}, {Ptak}, {Refsdal}, {Servillat}, \& {Streicher}}]{2013A&A...558A..33A}
{Astropy Collaboration}, {Robitaille}, T.~P., {Tollerud}, E.~J., {et~al.} 2013, \aap, 558, A33, \dodoi{10.1051/0004-6361/201322068}

\bibitem[{{Astropy Collaboration} {et~al.}(2018){Astropy Collaboration}, {Price-Whelan}, {Sip{\H{o}}cz}, {G{\"u}nther}, {Lim}, {Crawford}, {Conseil}, {Shupe}, {Craig}, {Dencheva}, {Ginsburg}, {VanderPlas}, {Bradley}, {P{\'e}rez-Su{\'a}rez}, {de Val-Borro}, {Aldcroft}, {Cruz}, {Robitaille}, {Tollerud}, {Ardelean}, {Babej}, {Bach}, {Bachetti}, {Bakanov}, {Bamford}, {Barentsen}, {Barmby}, {Baumbach}, {Berry}, {Biscani}, {Boquien}, {Bostroem}, {Bouma}, {Brammer}, {Bray}, {Breytenbach}, {Buddelmeijer}, {Burke}, {Calderone}, {Cano Rodr{\'\i}guez}, {Cara}, {Cardoso}, {Cheedella}, {Copin}, {Corrales}, {Crichton}, {D'Avella}, {Deil}, {Depagne}, {Dietrich}, {Donath}, {Droettboom}, {Earl}, {Erben}, {Fabbro}, {Ferreira}, {Finethy}, {Fox}, {Garrison}, {Gibbons}, {Goldstein}, {Gommers}, {Greco}, {Greenfield}, {Groener}, {Grollier}, {Hagen}, {Hirst}, {Homeier}, {Horton}, {Hosseinzadeh}, {Hu}, {Hunkeler}, {Ivezi{\'c}}, {Jain}, {Jenness}, {Kanarek}, {Kendrew}, {Kern}, {Kerzendorf}, {Khvalko}, {King}, {Kirkby}, {Kulkarni},
  {Kumar}, {Lee}, {Lenz}, {Littlefair}, {Ma}, {Macleod}, {Mastropietro}, {McCully}, {Montagnac}, {Morris}, {Mueller}, {Mumford}, {Muna}, {Murphy}, {Nelson}, {Nguyen}, {Ninan}, {N{\"o}the}, {Ogaz}, {Oh}, {Parejko}, {Parley}, {Pascual}, {Patil}, {Patil}, {Plunkett}, {Prochaska}, {Rastogi}, {Reddy Janga}, {Sabater}, {Sakurikar}, {Seifert}, {Sherbert}, {Sherwood-Taylor}, {Shih}, {Sick}, {Silbiger}, {Singanamalla}, {Singer}, {Sladen}, {Sooley}, {Sornarajah}, {Streicher}, {Teuben}, {Thomas}, {Tremblay}, {Turner}, {Terr{\'o}n}, {van Kerkwijk}, {de la Vega}, {Watkins}, {Weaver}, {Whitmore}, {Woillez}, {Zabalza}, \& {Astropy Contributors}}]{2018AJ....156..123A}
{Astropy Collaboration}, {Price-Whelan}, A.~M., {Sip{\H{o}}cz}, B.~M., {et~al.} 2018, \aj, 156, 123, \dodoi{10.3847/1538-3881/aabc4f}

\bibitem[{{Benisty} {et~al.}(2022){Benisty}, {Dominik}, {Follette}, {Garufi}, {Ginski}, {Hashimoto}, {Keppler}, {Kley}, \& {Monnier}}]{2022arXiv220309991B}
{Benisty}, M., {Dominik}, C., {Follette}, K., {et~al.} 2022, arXiv e-prints, arXiv:2203.09991, \dodoi{10.48550/arXiv.2203.09991}

\bibitem[{{Beuzit} {et~al.}(2019){Beuzit}, {Vigan}, {Mouillet}, {Dohlen}, {Gratton}, {Boccaletti}, {Sauvage}, {Schmid}, {Langlois}, {Petit}, {Baruffolo}, {Feldt}, {Milli}, {Wahhaj}, {Abe}, {Anselmi}, {Antichi}, {Barette}, {Baudrand}, {Baudoz}, {Bazzon}, {Bernardi}, {Blanchard}, {Brast}, {Bruno}, {Buey}, {Carbillet}, {Carle}, {Cascone}, {Chapron}, {Charton}, {Chauvin}, {Claudi}, {Costille}, {De Caprio}, {de Boer}, {Delboulb{\'e}}, {Desidera}, {Dominik}, {Downing}, {Dupuis}, {Fabron}, {Fantinel}, {Farisato}, {Feautrier}, {Fedrigo}, {Fusco}, {Gigan}, {Ginski}, {Girard}, {Giro}, {Gisler}, {Gluck}, {Gry}, {Henning}, {Hubin}, {Hugot}, {Incorvaia}, {Jaquet}, {Kasper}, {Lagadec}, {Lagrange}, {Le Coroller}, {Le Mignant}, {Le Ruyet}, {Lessio}, {Lizon}, {Llored}, {Lundin}, {Madec}, {Magnard}, {Marteaud}, {Martinez}, {Maurel}, {M{\'e}nard}, {Mesa}, {M{\"o}ller-Nilsson}, {Moulin}, {Moutou}, {Orign{\'e}}, {Parisot}, {Pavlov}, {Perret}, {Pragt}, {Puget}, {Rabou}, {Ramos}, {Reess}, {Rigal}, {Rochat}, {Roelfsema}, {Rousset},
  {Roux}, {Saisse}, {Salasnich}, {Santambrogio}, {Scuderi}, {Segransan}, {Sevin}, {Siebenmorgen}, {Soenke}, {Stadler}, {Suarez}, {Tiph{\`e}ne}, {Turatto}, {Udry}, {Vakili}, {Waters}, {Weber}, {Wildi}, {Zins}, \& {Zurlo}}]{2019A&A...631A.155B}
{Beuzit}, J.~L., {Vigan}, A., {Mouillet}, D., {et~al.} 2019, \aap, 631, A155, \dodoi{10.1051/0004-6361/201935251}

\bibitem[{{Blakely} {et~al.}(2022){Blakely}, {Francis}, {Johnstone}, {Soulain}, {Tuthill}, {Cheetham}, {Sanchez-Bermudez}, {Sivaramakrishnan}, {Dong}, {van der Marel}, {Cooper}, {Vigan}, \& {Cantalloube}}]{2022ApJ...931....3B}
{Blakely}, D., {Francis}, L., {Johnstone}, D., {et~al.} 2022, \apj, 931, 3, \dodoi{10.3847/1538-4357/ac6586}

\bibitem[{{Blakely} {et~al.}(2024){Blakely}, {Johnstone}, {Cugno}, {Sivaramakrishnan}, {Tuthill}, {Dong}, {Pope}, {Albert}, {Charles}, {Cooper}, {De Furio}, {Desdoigts}, {Doyon}, {Francis}, {Greenbaum}, {Lafreni{\`e}re}, {Lloyd}, {Meyer}, {Pueyo}, {Ray}, {S{\'a}nchez-Berm{\'u}dez}, {Soulain}, {Thatte}, {Thompson}, \& {Vandal}}]{2024arXiv240413032B}
{Blakely}, D., {Johnstone}, D., {Cugno}, G., {et~al.} 2024, arXiv e-prints, arXiv:2404.13032, \dodoi{10.48550/arXiv.2404.13032}

\bibitem[{{Boccaletti} {et~al.}(2013){Boccaletti}, {Pantin}, {Lagrange}, {Augereau}, {Meheut}, \& {Quanz}}]{2013A&A...560A..20B}
{Boccaletti}, A., {Pantin}, E., {Lagrange}, A.~M., {et~al.} 2013, \aap, 560, A20, \dodoi{10.1051/0004-6361/201322365}

\bibitem[{{Bohn} {et~al.}(2022){Bohn}, {Benisty}, {Perraut}, {van der Marel}, {W{\"o}lfer}, {van Dishoeck}, {Facchini}, {Manara}, {Teague}, {Francis}, {Berger}, {Garcia-Lopez}, {Ginski}, {Henning}, {Kenworthy}, {Kraus}, {M{\'e}nard}, {M{\'e}rand}, \& {P{\'e}rez}}]{2022AA...658A.183B}
{Bohn}, A.~J., {Benisty}, M., {Perraut}, K., {et~al.} 2022, \aap, 658, A183, \dodoi{10.1051/0004-6361/202142070}

\bibitem[{{Booth} {et~al.}(2023){Booth}, {Ilee}, {Walsh}, {Kama}, {Keyte}, {van Dishoeck}, \& {Nomura}}]{2023A&A...669A..53B}
{Booth}, A.~S., {Ilee}, J.~D., {Walsh}, C., {et~al.} 2023, \aap, 669, A53, \dodoi{10.1051/0004-6361/202244472}

\bibitem[{Bradbury {et~al.}(2018)Bradbury, Frostig, Hawkins, Johnson, Leary, Maclaurin, Necula, Paszke, Vander{P}las, Wanderman-{M}ilne, \& Zhang}]{jax2018github}
Bradbury, J., Frostig, R., Hawkins, P., {et~al.} 2018, {JAX}: composable transformations of {P}ython+{N}um{P}y programs, 0.3.13.
\newblock \url{http://github.com/google/jax}

\bibitem[{{Brittain} {et~al.}(2019){Brittain}, {Najita}, \& {Carr}}]{2019ApJ...883...37B}
{Brittain}, S.~D., {Najita}, J.~R., \& {Carr}, J.~S. 2019, \apj, 883, 37, \dodoi{10.3847/1538-4357/ab380b}

\bibitem[{{Casassus} {et~al.}(2022){Casassus}, {C{\'a}rcamo}, {Hales}, {Weber}, \& {Dent}}]{2022ApJ...933L...4C}
{Casassus}, S., {C{\'a}rcamo}, M., {Hales}, A., {Weber}, P., \& {Dent}, B. 2022, \apjl, 933, L4, \dodoi{10.3847/2041-8213/ac75e8}

\bibitem[{{Casassus} \& {P{\'e}rez}(2019)}]{2019ApJ...883L..41C}
{Casassus}, S., \& {P{\'e}rez}, S. 2019, \apjl, 883, L41, \dodoi{10.3847/2041-8213/ab4425}

\bibitem[{{Cheetham} {et~al.}(2016){Cheetham}, {Girard}, {Lacour}, {Schworer}, {Haubois}, \& {Beuzit}}]{2016SPIE.9907E..2TC}
{Cheetham}, A.~C., {Girard}, J., {Lacour}, S., {et~al.} 2016, in Society of Photo-Optical Instrumentation Engineers (SPIE) Conference Series, Vol. 9907, Optical and Infrared Interferometry and Imaging V, ed. F.~{Malbet}, M.~J. {Creech-Eakman}, \& P.~G. {Tuthill}, 99072T, \dodoi{10.1117/12.2231983}

\bibitem[{{Cieza} {et~al.}(2013){Cieza}, {Lacour}, {Schreiber}, {Casassus}, {Jord{\'a}n}, {Mathews}, {C{\'a}novas}, {M{\'e}nard}, {Kraus}, {P{\'e}rez}, {Tuthill}, \& {Ireland}}]{2013ApJ...762L..12C}
{Cieza}, L.~A., {Lacour}, S., {Schreiber}, M.~R., {et~al.} 2013, \apjl, 762, L12, \dodoi{10.1088/2041-8205/762/1/L12}

\bibitem[{{Currie} {et~al.}(2015){Currie}, {Cloutier}, {Brittain}, {Grady}, {Burrows}, {Muto}, {Kenyon}, \& {Kuchner}}]{2015ApJ...814L..27C}
{Currie}, T., {Cloutier}, R., {Brittain}, S., {et~al.} 2015, \apjl, 814, L27, \dodoi{10.1088/2041-8205/814/2/L27}

\bibitem[{{Currie} {et~al.}(2014){Currie}, {Muto}, {Kudo}, {Honda}, {Brandt}, {Grady}, {Fukagawa}, {Burrows}, {Janson}, {Kuzuhara}, {McElwain}, {Follette}, {Hashimoto}, {Henning}, {Kandori}, {Kusakabe}, {Kwon}, {Mede}, {Morino}, {Nishikawa}, {Pyo}, {Serabyn}, {Suenaga}, {Takahashi}, {Wisniewski}, \& {Tamura}}]{2014ApJ...796L..30C}
{Currie}, T., {Muto}, T., {Kudo}, T., {et~al.} 2014, \apjl, 796, L30, \dodoi{10.1088/2041-8205/796/2/L30}

\bibitem[{{Dohlen} {et~al.}(2008){Dohlen}, {Langlois}, {Saisse}, {Hill}, {Origne}, {Jacquet}, {Fabron}, {Blanc}, {Llored}, {Carle}, {Moutou}, {Vigan}, {Boccaletti}, {Carbillet}, {Mouillet}, \& {Beuzit}}]{2008SPIE.7014E..3LD}
{Dohlen}, K., {Langlois}, M., {Saisse}, M., {et~al.} 2008, in Society of Photo-Optical Instrumentation Engineers (SPIE) Conference Series, Vol. 7014, Ground-based and Airborne Instrumentation for Astronomy II, ed. I.~S. {McLean} \& M.~M. {Casali}, 70143L, \dodoi{10.1117/12.789786}

\bibitem[{{Fedele} {et~al.}(2021){Fedele}, {Toci}, {Maud}, \& {Lodato}}]{2021A&A...651A..90F}
{Fedele}, D., {Toci}, C., {Maud}, L., \& {Lodato}, G. 2021, \aap, 651, A90, \dodoi{10.1051/0004-6361/202141278}

\bibitem[{{Follette} {et~al.}(2017){Follette}, {Rameau}, {Dong}, {Pueyo}, {Close}, {Duch{\^e}ne}, {Fung}, {Leonard}, {Macintosh}, {Males}, {Marois}, {Millar-Blanchaer}, {Morzinski}, {Mullen}, {Perrin}, {Spiro}, {Wang}, {Ammons}, {Bailey}, {Barman}, {Bulger}, {Chilcote}, {Cotten}, {De Rosa}, {Doyon}, {Fitzgerald}, {Goodsell}, {Graham}, {Greenbaum}, {Hibon}, {Hung}, {Ingraham}, {Kalas}, {Konopacky}, {Larkin}, {Maire}, {Marchis}, {Metchev}, {Nielsen}, {Oppenheimer}, {Palmer}, {Patience}, {Poyneer}, {Rajan}, {Rantakyr{\"o}}, {Savransky}, {Schneider}, {Sivaramakrishnan}, {Song}, {Soummer}, {Thomas}, {Vega}, {Wallace}, {Ward-Duong}, {Wiktorowicz}, \& {Wolff}}]{2017AJ....153..264F}
{Follette}, K.~B., {Rameau}, J., {Dong}, R., {et~al.} 2017, \aj, 153, 264, \dodoi{10.3847/1538-3881/aa6d85}

\bibitem[{Foreman-Mackey(2016)}]{corner}
Foreman-Mackey, D. 2016, The Journal of Open Source Software, 1, 24, \dodoi{10.21105/joss.00024}

\bibitem[{{Gaia Collaboration} {et~al.}(2023){Gaia Collaboration}, {Vallenari}, {Brown}, {Prusti}, {de Bruijne}, {Arenou}, {Babusiaux}, {Biermann}, {Creevey}, {Ducourant}, {Evans}, {Eyer}, {Guerra}, {Hutton}, {Jordi}, {Klioner}, {Lammers}, {Lindegren}, {Luri}, {Mignard}, {Panem}, {Pourbaix}, {Randich}, {Sartoretti}, {Soubiran}, {Tanga}, {Walton}, {Bailer-Jones}, {Bastian}, {Drimmel}, {Jansen}, {Katz}, {Lattanzi}, {van Leeuwen}, {Bakker}, {Cacciari}, {Casta{\~n}eda}, {De Angeli}, {Fabricius}, {Fouesneau}, {Fr{\'e}mat}, {Galluccio}, {Guerrier}, {Heiter}, {Masana}, {Messineo}, {Mowlavi}, {Nicolas}, {Nienartowicz}, {Pailler}, {Panuzzo}, {Riclet}, {Roux}, {Seabroke}, {Sordo}, {Th{\'e}venin}, {Gracia-Abril}, {Portell}, {Teyssier}, {Altmann}, {Andrae}, {Audard}, {Bellas-Velidis}, {Benson}, {Berthier}, {Blomme}, {Burgess}, {Busonero}, {Busso}, {C{\'a}novas}, {Carry}, {Cellino}, {Cheek}, {Clementini}, {Damerdji}, {Davidson}, {de Teodoro}, {Nu{\~n}ez Campos}, {Delchambre}, {Dell'Oro}, {Esquej},
  {Fern{\'a}ndez-Hern{\'a}ndez}, {Fraile}, {Garabato}, {Garc{\'\i}a-Lario}, {Gosset}, {Haigron}, {Halbwachs}, {Hambly}, {Harrison}, {Hern{\'a}ndez}, {Hestroffer}, {Hodgkin}, {Holl}, {Jan{\ss}en}, {Jevardat de Fombelle}, {Jordan}, {Krone-Martins}, {Lanzafame}, {L{\"o}ffler}, {Marchal}, {Marrese}, {Moitinho}, {Muinonen}, {Osborne}, {Pancino}, {Pauwels}, {Recio-Blanco}, {Reyl{\'e}}, {Riello}, {Rimoldini}, {Roegiers}, {Rybizki}, {Sarro}, {Siopis}, {Smith}, {Sozzetti}, {Utrilla}, {van Leeuwen}, {Abbas}, {{\'A}brah{\'a}m}, {Abreu Aramburu}, {Aerts}, {Aguado}, {Ajaj}, {Aldea-Montero}, {Altavilla}, {{\'A}lvarez}, {Alves}, {Anders}, {Anderson}, {Anglada Varela}, {Antoja}, {Baines}, {Baker}, {Balaguer-N{\'u}{\~n}ez}, {Balbinot}, {Balog}, {Barache}, {Barbato}, {Barros}, {Barstow}, {Bartolom{\'e}}, {Bassilana}, {Bauchet}, {Becciani}, {Bellazzini}, {Berihuete}, {Bernet}, {Bertone}, {Bianchi}, {Binnenfeld}, {Blanco-Cuaresma}, {Blazere}, {Boch}, {Bombrun}, {Bossini}, {Bouquillon}, {Bragaglia}, {Bramante}, {Breedt},
  {Bressan}, {Brouillet}, {Brugaletta}, {Bucciarelli}, {Burlacu}, {Butkevich}, {Buzzi}, {Caffau}, {Cancelliere}, {Cantat-Gaudin}, {Carballo}, {Carlucci}, {Carnerero}, {Carrasco}, {Casamiquela}, {Castellani}, {Castro-Ginard}, {Chaoul}, {Charlot}, {Chemin}, {Chiaramida}, {Chiavassa}, {Chornay}, {Comoretto}, {Contursi}, {Cooper}, {Cornez}, {Cowell}, {Crifo}, {Cropper}, {Crosta}, {Crowley}, {Dafonte}, {Dapergolas}, {David}, {David}, {de Laverny}, {De Luise}, {De March}, {De Ridder}, {de Souza}, {de Torres}, {del Peloso}, {del Pozo}, {Delbo}, {Delgado}, {Delisle}, {Demouchy}, {Dharmawardena}, {Di Matteo}, {Diakite}, {Diener}, {Distefano}, {Dolding}, {Edvardsson}, {Enke}, {Fabre}, {Fabrizio}, {Faigler}, {Fedorets}, {Fernique}, {Fienga}, {Figueras}, {Fournier}, {Fouron}, {Fragkoudi}, {Gai}, {Garcia-Gutierrez}, {Garcia-Reinaldos}, {Garc{\'\i}a-Torres}, {Garofalo}, {Gavel}, {Gavras}, {Gerlach}, {Geyer}, {Giacobbe}, {Gilmore}, {Girona}, {Giuffrida}, {Gomel}, {Gomez}, {Gonz{\'a}lez-N{\'u}{\~n}ez},
  {Gonz{\'a}lez-Santamar{\'\i}a}, {Gonz{\'a}lez-Vidal}, {Granvik}, {Guillout}, {Guiraud}, {Guti{\'e}rrez-S{\'a}nchez}, {Guy}, {Hatzidimitriou}, {Hauser}, {Haywood}, {Helmer}, {Helmi}, {Sarmiento}, {Hidalgo}, {Hilger}, {H{\l}adczuk}, {Hobbs}, {Holland}, {Huckle}, {Jardine}, {Jasniewicz}, {Jean-Antoine Piccolo}, {Jim{\'e}nez-Arranz}, {Jorissen}, {Juaristi Campillo}, {Julbe}, {Karbevska}, {Kervella}, {Khanna}, {Kontizas}, {Kordopatis}, {Korn}, {K{\'o}sp{\'a}l}, {Kostrzewa-Rutkowska}, {Kruszy{\'n}ska}, {Kun}, {Laizeau}, {Lambert}, {Lanza}, {Lasne}, {Le Campion}, {Lebreton}, {Lebzelter}, {Leccia}, {Leclerc}, {Lecoeur-Taibi}, {Liao}, {Licata}, {Lindstr{\o}m}, {Lister}, {Livanou}, {Lobel}, {Lorca}, {Loup}, {Madrero Pardo}, {Magdaleno Romeo}, {Managau}, {Mann}, {Manteiga}, {Marchant}, {Marconi}, {Marcos}, {Marcos Santos}, {Mar{\'\i}n Pina}, {Marinoni}, {Marocco}, {Marshall}, {Martin Polo}, {Mart{\'\i}n-Fleitas}, {Marton}, {Mary}, {Masip}, {Massari}, {Mastrobuono-Battisti}, {Mazeh}, {McMillan}, {Messina}, {Michalik},
  {Millar}, {Mints}, {Molina}, {Molinaro}, {Moln{\'a}r}, {Monari}, {Mongui{\'o}}, {Montegriffo}, {Montero}, {Mor}, {Mora}, {Morbidelli}, {Morel}, {Morris}, {Muraveva}, {Murphy}, {Musella}, {Nagy}, {Noval}, {Oca{\~n}a}, {Ogden}, {Ordenovic}, {Osinde}, {Pagani}, {Pagano}, {Palaversa}, {Palicio}, {Pallas-Quintela}, {Panahi}, {Payne-Wardenaar}, {Pe{\~n}alosa Esteller}, {Penttil{\"a}}, {Pichon}, {Piersimoni}, {Pineau}, {Plachy}, {Plum}, {Poggio}, {Pr{\v{s}}a}, {Pulone}, {Racero}, {Ragaini}, {Rainer}, {Raiteri}, {Rambaux}, {Ramos}, {Ramos-Lerate}, {Re Fiorentin}, {Regibo}, {Richards}, {Rios Diaz}, {Ripepi}, {Riva}, {Rix}, {Rixon}, {Robichon}, {Robin}, {Robin}, {Roelens}, {Rogues}, {Rohrbasser}, {Romero-G{\'o}mez}, {Rowell}, {Royer}, {Ruz Mieres}, {Rybicki}, {Sadowski}, {S{\'a}ez N{\'u}{\~n}ez}, {Sagrist{\`a} Sell{\'e}s}, {Sahlmann}, {Salguero}, {Samaras}, {Sanchez Gimenez}, {Sanna}, {Santove{\~n}a}, {Sarasso}, {Schultheis}, {Sciacca}, {Segol}, {Segovia}, {S{\'e}gransan}, {Semeux}, {Shahaf}, {Siddiqui}, {Siebert},
  {Siltala}, {Silvelo}, {Slezak}, {Slezak}, {Smart}, {Snaith}, {Solano}, {Solitro}, {Souami}, {Souchay}, {Spagna}, {Spina}, {Spoto}, {Steele}, {Steidelm{\"u}ller}, {Stephenson}, {S{\"u}veges}, {Surdej}, {Szabados}, {Szegedi-Elek}, {Taris}, {Taylor}, {Teixeira}, {Tolomei}, {Tonello}, {Torra}, {Torra}, {Torralba Elipe}, {Trabucchi}, {Tsounis}, {Turon}, {Ulla}, {Unger}, {Vaillant}, {van Dillen}, {van Reeven}, {Vanel}, {Vecchiato}, {Viala}, {Vicente}, {Voutsinas}, {Weiler}, {Wevers}, {Wyrzykowski}, {Yoldas}, {Yvard}, {Zhao}, {Zorec}, {Zucker}, \& {Zwitter}}]{2023AA...674A...1G}
{Gaia Collaboration}, {Vallenari}, A., {Brown}, A.~G.~A., {et~al.} 2023, \aap, 674, A1, \dodoi{10.1051/0004-6361/202243940}

\bibitem[{{Gallenne} {et~al.}(2015){Gallenne}, {M{\'e}rand}, {Kervella}, {Monnier}, {Schaefer}, {Baron}, {Breitfelder}, {Le Bouquin}, {Roettenbacher}, {Gieren}, {Pietrzy{\'n}ski}, {McAlister}, {ten Brummelaar}, {Sturmann}, {Sturmann}, {Turner}, {Ridgway}, \& {Kraus}}]{2015A&A...579A..68G}
{Gallenne}, A., {M{\'e}rand}, A., {Kervella}, P., {et~al.} 2015, \aap, 579, A68, \dodoi{10.1051/0004-6361/201525917}

\bibitem[{{Garufi} {et~al.}(2016){Garufi}, {Quanz}, {Schmid}, {Mulders}, {Avenhaus}, {Boccaletti}, {Ginski}, {Langlois}, {Stolker}, {Augereau}, {Benisty}, {Lopez}, {Dominik}, {Gratton}, {Henning}, {Janson}, {M{\'e}nard}, {Meyer}, {Pinte}, {Sissa}, {Vigan}, {Zurlo}, {Bazzon}, {Buenzli}, {Bonnefoy}, {Brandner}, {Chauvin}, {Cheetham}, {Cudel}, {Desidera}, {Feldt}, {Galicher}, {Kasper}, {Lagrange}, {Lannier}, {Maire}, {Mesa}, {Mouillet}, {Peretti}, {Perrot}, {Salter}, \& {Wildi}}]{2016A&A...588A...8G}
{Garufi}, A., {Quanz}, S.~P., {Schmid}, H.~M., {et~al.} 2016, \aap, 588, A8, \dodoi{10.1051/0004-6361/201527940}

\bibitem[{Harris {et~al.}(2020)Harris, Millman, van~der Walt, Gommers, Virtanen, Cournapeau, Wieser, Taylor, Berg, Smith, Kern, Picus, Hoyer, van Kerkwijk, Brett, Haldane, del R{\'{i}}o, Wiebe, Peterson, G{\'{e}}rard-Marchant, Sheppard, Reddy, Weckesser, Abbasi, Gohlke, \& Oliphant}]{harris2020array}
Harris, C.~R., Millman, K.~J., van~der Walt, S.~J., {et~al.} 2020, Nature, 585, 357, \dodoi{10.1038/s41586-020-2649-2}

\bibitem[{Higson {et~al.}(2018)Higson, Handley, Hobson, \& Lasenby}]{dyn2018}
Higson, E., Handley, W., Hobson, M., \& Lasenby, A. 2018, Statistics and Computing, 29, 891–913, \dodoi{10.1007/s11222-018-9844-0}

\bibitem[{Hunter(2007)}]{Hunter:2007}
Hunter, J.~D. 2007, Computing in Science \& Engineering, 9, 90, \dodoi{10.1109/MCSE.2007.55}

\bibitem[{{Jamialahmadi} {et~al.}(2018){Jamialahmadi}, {Ratzka}, {Pani{\'c}}, {Fathivavsari}, {van Boekel}, {Flement}, {Henning}, {Jaffe}, \& {Mulders}}]{2018ApJ...865..137J}
{Jamialahmadi}, N., {Ratzka}, T., {Pani{\'c}}, O., {et~al.} 2018, \apj, 865, 137, \dodoi{10.3847/1538-4357/aadae4}

\bibitem[{Jenkins \& Peacock(2011)}]{bayesevid2011}
Jenkins, C.~R., \& Peacock, J.~A. 2011, Monthly Notices of the Royal Astronomical Society, 413, 2895–2905, \dodoi{10.1111/j.1365-2966.2011.18361.x}

\bibitem[{Kass \& Raftery(1995)}]{doi:10.1080/01621459.1995.10476572}
Kass, R.~E., \& Raftery, A.~E. 1995, Journal of the American Statistical Association, 90, 773, \dodoi{10.1080/01621459.1995.10476572}

\bibitem[{{Keyte} {et~al.}(2023){Keyte}, {Kama}, {Booth}, {Bergin}, {Cleeves}, {van Dishoeck}, {Drozdovskaya}, {Furuya}, {Rawlings}, {Shorttle}, \& {Walsh}}]{2023NatAs...7..684K}
{Keyte}, L., {Kama}, M., {Booth}, A.~S., {et~al.} 2023, Nature Astronomy, 7, 684, \dodoi{10.1038/s41550-023-01951-9}

\bibitem[{Koposov {et~al.}(2023)Koposov, Speagle, Barbary, Ashton, Bennett, Buchner, Scheffler, Cook, Talbot, Guillochon, Cubillos, Ramos, Johnson, Lang, Ilya, Dartiailh, Nitz, McCluskey, Archibald, Deil, Foreman-Mackey, Goldstein, Tollerud, Leja, Kirk, Pitkin, Sheehan, Cargile, Patel, \& Angus}]{sergey_koposov_2023_7600689}
Koposov, S., Speagle, J., Barbary, K., {et~al.} 2023, joshspeagle/dynesty: v2.1.0, v2.1.0,  Zenodo, \dodoi{10.5281/zenodo.7600689}

\bibitem[{{Lacour} {et~al.}(2011){Lacour}, {Tuthill}, {Ireland}, {Amico}, \& {Girard}}]{2011Msngr.146...18L}
{Lacour}, S., {Tuthill}, P., {Ireland}, M., {Amico}, P., \& {Girard}, J. 2011, The Messenger, 146, 18

\bibitem[{{Lazareff} {et~al.}(2017){Lazareff}, {Berger}, {Kluska}, {Le Bouquin}, {Benisty}, {Malbet}, {Koen}, {Pinte}, {Thi}, {Absil}, {Baron}, {Delboulb{\'e}}, {Duvert}, {Isella}, {Jocou}, {Juhasz}, {Kraus}, {Lachaume}, {M{\'e}nard}, {Millan-Gabet}, {Monnier}, {Moulin}, {Perraut}, {Rochat}, {Soulez}, {Tallon}, {Thi{\'e}baut}, {Traub}, \& {Zins}}]{2017AA...599A..85L}
{Lazareff}, B., {Berger}, J.~P., {Kluska}, J., {et~al.} 2017, \aap, 599, A85, \dodoi{10.1051/0004-6361/201629305}

\bibitem[{{Miley} {et~al.}(2019){Miley}, {Pani{\'c}}, {Haworth}, {Pascucci}, {Wyatt}, {Clarke}, {Richards}, \& {Ratzka}}]{2019MNRAS.485..739M}
{Miley}, J.~M., {Pani{\'c}}, O., {Haworth}, T.~J., {et~al.} 2019, \mnras, 485, 739, \dodoi{10.1093/mnras/stz426}

\bibitem[{{Norfolk} {et~al.}(2021){Norfolk}, {Maddison}, {Pinte}, {van der Marel}, {Booth}, {Francis}, {Gonzalez}, {M{\'e}nard}, {Wright}, {van der Plas}, \& {Garg}}]{2021MNRAS.502.5779N}
{Norfolk}, B.~J., {Maddison}, S.~T., {Pinte}, C., {et~al.} 2021, \mnras, 502, 5779, \dodoi{10.1093/mnras/stab313}

\bibitem[{{Norfolk} {et~al.}(2022){Norfolk}, {Pinte}, {Calcino}, {Hammond}, {van der Marel}, {Price}, {Maddison}, {Christiaens}, {Gonzalez}, {Blakely}, {Rosotti}, \& {Ginski}}]{2022ApJ...936L...4N}
{Norfolk}, B.~J., {Pinte}, C., {Calcino}, J., {et~al.} 2022, \apjl, 936, L4, \dodoi{10.3847/2041-8213/ac85ed}

\bibitem[{{Pani{\'c}} {et~al.}(2014){Pani{\'c}}, {Ratzka}, {Mulders}, {Dominik}, {van Boekel}, {Henning}, {Jaffe}, \& {Min}}]{2014A&A...562A.101P}
{Pani{\'c}}, O., {Ratzka}, T., {Mulders}, G.~D., {et~al.} 2014, \aap, 562, A101, \dodoi{10.1051/0004-6361/201219223}

\bibitem[{{P{\'e}rez} {et~al.}(2020){P{\'e}rez}, {Casassus}, {Hales}, {Marino}, {Cheetham}, {Zurlo}, {Cieza}, {Dong}, {Alarc{\'o}n}, {Ben{\'\i}tez-Llambay}, {Fomalont}, \& {Avenhaus}}]{2020ApJ...889L..24P}
{P{\'e}rez}, S., {Casassus}, S., {Hales}, A., {et~al.} 2020, \apjl, 889, L24, \dodoi{10.3847/2041-8213/ab6b2b}

\bibitem[{{Phillips} {et~al.}(2020){Phillips}, {Tremblin}, {Baraffe}, {Chabrier}, {Allard}, {Spiegelman}, {Goyal}, {Drummond}, \& {H{\'e}brard}}]{2020A&A...637A..38P}
{Phillips}, M.~W., {Tremblin}, P., {Baraffe}, I., {et~al.} 2020, \aap, 637, A38, \dodoi{10.1051/0004-6361/201937381}

\bibitem[{{Pineda} {et~al.}(2019){Pineda}, {Szul{\'a}gyi}, {Quanz}, {van Dishoeck}, {Garufi}, {Meru}, {Mulders}, {Testi}, {Meyer}, \& {Reggiani}}]{2019ApJ...871...48P}
{Pineda}, J.~E., {Szul{\'a}gyi}, J., {Quanz}, S.~P., {et~al.} 2019, \apj, 871, 48, \dodoi{10.3847/1538-4357/aaf389}

\bibitem[{{Pourr{\'e}} {et~al.}(2024){Pourr{\'e}}, {Winterhalder}, {Le Bouquin}, {Lacour}, {Bidot}, {Nowak}, {Maire}, {Mouillet}, {Babusiaux}, {Woillez}, {Abuter}, {Amorim}, {Asensio-Torres}, {Balmer}, {Benisty}, {Berger}, {Beust}, {Blunt}, {Boccaletti}, {Bonnefoy}, {Bonnet}, {Bordoni}, {Bourdarot}, {Brandner}, {Cantalloube}, {Caselli}, {Charnay}, {Chauvin}, {Chavez}, {Choquet}, {Christiaens}, {Cl{\'e}net}, {du Foresto}, {Cridland}, {Davies}, {Defr{\`e}re}, {Dembet}, {Dexter}, {Drescher}, {Duvert}, {Eckart}, {Eisenhauer}, {Schreiber}, {Garcia}, {Lopez}, {Gendron}, {Genzel}, {Gillessen}, {Girard}, {Gonte}, {Grant}, {Haubois}, {Hei{\ss}el}, {Henning}, {Hinkley}, {Hippler}, {H{\"o}nig}, {Houll{\'e}}, {Hubert}, {Jocou}, {Kammerer}, {Kenworthy}, {Keppler}, {Kervella}, {Kreidberg}, {Kurtovic}, {Lagrange}, {Lapeyr{\`e}re}, {Lutz}, {Mang}, {Marleau}, {M{\'e}rand}, {Millour}, {Molli{\`e}re}, {Monnier}, {Mordasini}, {Nasedkin}, {Oberti}, {Ott}, {Otten}, {Paladini}, {Paumard}, {Perraut}, {Perrin}, {Pfuhl}, {Pueyo},
  {Ribeiro}, {Rickman}, {Rustamkulov}, {Shangguan}, {Shimizu}, {Sing}, {Soulez}, {Stadler}, {Stolker}, {Straub}, {Straubmeier}, {Sturm}, {Sykes}, {Tacconi}, {van Dishoeck}, {Vigan}, {Vincent}, {von Fellenberg}, {Wang}, {Widmann}, {Yazici}, {Abad}, {Carpentier}, {Alonso}, {Andolfato}, {Barriga}, {Beuzit}, {Bourget}, {Brast}, {Caniguante}, {Cottalorda}, {Darr{\'e}}, {Delabre}, {Delboulb{\'e}}, {Delplancke-Str{\"o}bele}, {Donaldson}, {Dorn}, {Dupuy}, {Egner}, {Fischer}, {Frank}, {Fuenteseca}, {Gitton}, {Guerlet}, {Guieu}, {Gutierrez}, {Haguenauer}, {Haimerl}, {Heritier}, {Huber}, {Hubin}, {Jolley}, {Kirchbauer}, {Kolb}, {Kosmalski}, {Krempl}, {Le Louarn}, {Lilley}, {Lopez}, {Magnard}, {Mclay}, {Meilland}, {Meister}, {Moulin}, {Pasquini}, {Paufique}, {Percheron}, {Pettazzi}, {Phan}, {Pirani}, {Quentin}, {Rakich}, {Ridings}, {Reyes}, {Rochat}, {Schmid}, {Schuhler}, {Shchekaturov}, {Seidel}, {Soenke}, {Stadler}, {Stephan}, {Su{\'a}rez}, {Todorovic}, {Valdes}, {Verinaud}, {Zins}, \&
  {Z{\'u}{\~n}iga-Fern{\'a}ndez}}]{2024A&A...686A.258P}
{Pourr{\'e}}, N., {Winterhalder}, T.~O., {Le Bouquin}, J.~B., {et~al.} 2024, \aap, 686, A258, \dodoi{10.1051/0004-6361/202449507}

\bibitem[{{Quanz} {et~al.}(2013){Quanz}, {Amara}, {Meyer}, {Kenworthy}, {Kasper}, \& {Girard}}]{2013ApJ...766L...1Q}
{Quanz}, S.~P., {Amara}, A., {Meyer}, M.~R., {et~al.} 2013, \apjl, 766, L1, \dodoi{10.1088/2041-8205/766/1/L1}

\bibitem[{{Quanz} {et~al.}(2011){Quanz}, {Schmid}, {Geissler}, {Meyer}, {Henning}, {Brandner}, \& {Wolf}}]{2011ApJ...738...23Q}
{Quanz}, S.~P., {Schmid}, H.~M., {Geissler}, K., {et~al.} 2011, \apj, 738, 23, \dodoi{10.1088/0004-637X/738/1/23}

\bibitem[{{Ren} {et~al.}(2023){Ren}, {Benisty}, {Ginski}, {Tazaki}, {Wallack}, {Milli}, {Garufi}, {Bae}, {Facchini}, {M{\'e}nard}, {Pinilla}, {Swastik}, {Teague}, \& {Wahhaj}}]{2023A&A...680A.114R}
{Ren}, B.~B., {Benisty}, M., {Ginski}, C., {et~al.} 2023, \aap, 680, A114, \dodoi{10.1051/0004-6361/202347353}

\bibitem[{{Sallum} {et~al.}(2023){Sallum}, {Eisner}, {Skemer}, \& {Murray-Clay}}]{2023ApJ...953...55S}
{Sallum}, S., {Eisner}, J., {Skemer}, A., \& {Murray-Clay}, R. 2023, \apj, 953, 55, \dodoi{10.3847/1538-4357/ace16c}

\bibitem[{{Skrutskie} {et~al.}(2006){Skrutskie}, {Cutri}, {Stiening}, {Weinberg}, {Schneider}, {Carpenter}, {Beichman}, {Capps}, {Chester}, {Elias}, {Huchra}, {Liebert}, {Lonsdale}, {Monet}, {Price}, {Seitzer}, {Jarrett}, {Kirkpatrick}, {Gizis}, {Howard}, {Evans}, {Fowler}, {Fullmer}, {Hurt}, {Light}, {Kopan}, {Marsh}, {McCallon}, {Tam}, {Van Dyk}, \& {Wheelock}}]{2006AJ....131.1163S}
{Skrutskie}, M.~F., {Cutri}, R.~M., {Stiening}, R., {et~al.} 2006, \aj, 131, 1163, \dodoi{10.1086/498708}

\bibitem[{{Soulain} {et~al.}(2020){Soulain}, {Sivaramakrishnan}, {Tuthill}, {Thatte}, {Volk}, {Cooper}, {Albert}, {Artigau}, {Cook}, {Doyon}, {Johnstone}, {Lafreni{\`e}re}, \& {Martel}}]{2020SPIE11446E..11S}
{Soulain}, A., {Sivaramakrishnan}, A., {Tuthill}, P., {et~al.} 2020, in Society of Photo-Optical Instrumentation Engineers (SPIE) Conference Series, Vol. 11446, Optical and Infrared Interferometry and Imaging VII, ed. P.~G. {Tuthill}, A.~{M{\'e}rand}, \& S.~{Sallum}, 1144611, \dodoi{10.1117/12.2560804}

\bibitem[{{Speagle}(2020)}]{2020MNRAS.493.3132S}
{Speagle}, J.~S. 2020, \mnras, 493, 3132, \dodoi{10.1093/mnras/staa278}

\bibitem[{{Stolker} {et~al.}(2020){Stolker}, {Quanz}, {Todorov}, {K{\"u}hn}, {Molli{\`e}re}, {Meyer}, {Currie}, {Daemgen}, \& {Lavie}}]{2020A&A...635A.182S}
{Stolker}, T., {Quanz}, S.~P., {Todorov}, K.~O., {et~al.} 2020, \aap, 635, A182, \dodoi{10.1051/0004-6361/201937159}

\bibitem[{{Stolker} {et~al.}(2024){Stolker}, {Kammerer}, {Benisty}, {Blakely}, {Johnstone}, {Sitko}, {Berger}, {Sanchez-Bermudez}, {Garufi}, {Lacour}, {Cantalloube}, \& {Chauvin}}]{2024A&A...682A.101S}
{Stolker}, T., {Kammerer}, J., {Benisty}, M., {et~al.} 2024, \aap, 682, A101, \dodoi{10.1051/0004-6361/202347291}

\bibitem[{Surjanovic {et~al.}(2023)Surjanovic, Biron-Lattes, Tiede, Syed, Campbell, \& Bouchard-C{\^o}t{\'e}}]{surjanovic2023pigeons}
Surjanovic, N., Biron-Lattes, M., Tiede, P., {et~al.} 2023, arXiv:2308.09769

\bibitem[{Syed {et~al.}(2022)Syed, Bouchard-Côté, Deligiannidis, \& Doucet}]{https://doi.org/10.1111/rssb.12464}
Syed, S., Bouchard-Côté, A., Deligiannidis, G., \& Doucet, A. 2022, Journal of the Royal Statistical Society: Series B (Statistical Methodology), 84, 321, \dodoi{https://doi.org/10.1111/rssb.12464}

\bibitem[{{Tatulli} {et~al.}(2011){Tatulli}, {Benisty}, {M{\'e}nard}, {Varni{\`e}re}, {Martin-Za{\"\i}di}, {Thi}, {Pinte}, {Massi}, {Weigelt}, {Hofmann}, \& {Petrov}}]{2011A&A...531A...1T}
{Tatulli}, E., {Benisty}, M., {M{\'e}nard}, F., {et~al.} 2011, \aap, 531, A1, \dodoi{10.1051/0004-6361/201016165}

\bibitem[{{Thompson} {et~al.}(2023){Thompson}, {Lawrence}, {Blakely}, {Marois}, {Wang}, {Giordano}, {Brandt}, {Johnstone}, {Ruffio}, {Ammons}, {Crotts}, {Do {\'O}}, {Gonzales}, \& {Rice}}]{2023AJ....166..164T}
{Thompson}, W., {Lawrence}, J., {Blakely}, D., {et~al.} 2023, \aj, 166, 164, \dodoi{10.3847/1538-3881/acf5cc}

\bibitem[{{Tuthill} {et~al.}(2010){Tuthill}, {Lacour}, {Amico}, {Ireland}, {Norris}, {Stewart}, {Evans}, {Kraus}, {Lidman}, {Pompei}, \& {Kornweibel}}]{2010SPIE.7735E..1OT}
{Tuthill}, P., {Lacour}, S., {Amico}, P., {et~al.} 2010, in Society of Photo-Optical Instrumentation Engineers (SPIE) Conference Series, Vol. 7735, Ground-based and Airborne Instrumentation for Astronomy III, ed. I.~S. {McLean}, S.~K. {Ramsay}, \& H.~{Takami}, 77351O, \dodoi{10.1117/12.856806}

\bibitem[{{Vigan}(2020)}]{2020ascl.soft09002V}
{Vigan}, A. 2020, {vlt-sphere: Automatic VLT/SPHERE data reduction and analysis}, Astrophysics Source Code Library, record ascl:2009.002.
\newblock \doeprint{2009.002}

\bibitem[{{Vigan} {et~al.}(2010){Vigan}, {Moutou}, {Langlois}, {Allard}, {Boccaletti}, {Carbillet}, {Mouillet}, \& {Smith}}]{2010MNRAS.407...71V}
{Vigan}, A., {Moutou}, C., {Langlois}, M., {et~al.} 2010, \mnras, 407, 71, \dodoi{10.1111/j.1365-2966.2010.16916.x}

\bibitem[{{Wright} {et~al.}(2015){Wright}, {Maddison}, {Wilner}, {Burton}, {Lommen}, {van Dishoeck}, {Pinilla}, {Bourke}, {Menard}, \& {Walsh}}]{2015MNRAS.453..414W}
{Wright}, C.~M., {Maddison}, S.~T., {Wilner}, D.~J., {et~al.} 2015, \mnras, 453, 414, \dodoi{10.1093/mnras/stv1619}

\end{thebibliography}
\bibliographystyle{aasjournal}



\end{document}